\documentclass[useAMS,usegraphicx,usenatbib]{mn2e}
\usepackage{multirow}
\usepackage{amssymb}

\newcommand{\kms}{\mbox{$\>{\rm km\, s^{-1}}$}}
\newcommand{\pc}{\>{\rm pc}}
\newcommand{\kpc}{\mbox{$\>{\rm kpc}$}} 
\newcommand{\Gyr}{\mbox{$\>{\rm Gyr}$}}
\newcommand{\msun}{\>{\rm M_{\odot}}}
\newcommand\degrees{^\circ}
 
\def\etal{{et al.}}

\def\ie{{\it i.e.}}

\title[The Milky Way's Disc Relative to the Halo]{What's Up in the
  Milky Way?  The Orientation of the Disc Relative to the Triaxial
  Halo}
   
\author[Debattista \etal]{Victor P. Debattista$^{1,2}$\thanks{E-mail:
    vpdebattista@gmail.com}\thanks{Visiting Lecturer, Department of Physics, University of Malta}, Rok Ro\v{s}kar$^3$, Monica Valluri$^4$,
  Thomas Quinn$^5$, \newauthor Ben Moore$^3$, James Wadsley$^6$ \\
$^1$ Jeremiah Horrocks Institute, University of Central Lancashire,
Preston, PR1 2HE, UK \\ 
$^2$ Department of Physics, University of Malta,
Tal-Qroqq Street, Msida, MSD 2080, Malta \\
$^3$ Institute of Theoretical Physics, University of Z\"urich,
Winterthurerstrasse 190, CH-8057, Z\"urich, Switzerland \\
$^4$ Department of Astronomy, University of Michigan, Ann Arbor, MI
48109, USA \\
$^5$ Astronomy Department, University of Washington, Box 351580,
Seattle, WA 98195, USA \\
$^6$ Department of Physics and Astronomy, McMaster University,
Hamilton, ON L8S 4M1, Canada 
}

\begin{document}   

\date{{\it Draft version on \today}}
\pagerange{\pageref{firstpage}--\pageref{lastpage}} \pubyear{----}
\maketitle

\label{firstpage}

\begin{abstract} 
  Models of the Sagittarius Stream have consistently found that the
  Milky Way disc is oriented such that its short axis is along the
  intermediate axis of the triaxial dark matter halo.  We attempt to
  build models of disc galaxies in such an `intermediate-axis
  orientation'.  We do this with three models.  In the first two cases
  we simply rigidly grow a disc in a triaxial halo such that the disc
  ends up perpendicular to the global intermediate axis.  We also
  attempt to coax a disc to form in an intermediate-axis orientation
  by producing a gas+dark matter triaxial system with gas angular
  momentum about the intermediate axis.  In all cases we fail to
  produce systems which remain with stellar angular momentum aligned
  with the halo's intermediate axis, even when the disc's potential
  flattens the inner halo such that the disc is everywhere
  perpendicular to the halo's local minor axis.  For one of these
  unstable simulations we show that the potential is even rounder than
  the models of the Milky Way potential in the region probed by the
  Sagittarius Stream.  We conclude that the Milky Way's disc is very
  unlikely to be in an intermediate axis orientation.  However we find
  that a disc can persist off one of the principal planes of the
  potential.  We propose that the disc of the Milky Way must be tilted
  relative to the principal axes of the dark matter halo.  Direct
  confirmation of this prediction would constitute a critical test of
  Modified Newtonian Dynamics.
\end{abstract}


%
%

\section{Introduction}
\label{sec:intro}

Dark matter haloes in purely collisionless simulations are generally
triaxial \citep{bbks_86, bar_efs_87, frenk_etal_88, dub_car_91,
  jin_sut_02, bai_ste_05, all_etal_06} with typical axial ratios $b/a
\sim 0.6$ and $c/a \sim 0.4$ (where $c < b < a$ are the short,
intermediate and long axes, respectively).  Observations instead find
rounder haloes \citep{sch_etal_83, sac_spa_90, fra_dez_92, hui_van_92,
  buote_canizares94, kui_tre_94, fra_etal_94, olling_95, kochan_95,
  bar_etal_95, olling_96, sch_etal_97, koo_etal_98, oll_mer_00,
  and_etal_01, buo_etal_02, ogu_etal_03, iod_etal_03, debatt_03,
  bar_sel_03, die_sta_07, spekkens_sellwood07, ban_jog_08}.  This
discrepancy is most likely accounted for by the fact that haloes become
rounder when baryons condense within them \citep{dubins_94, kkzanm04,
  debattista+08, zemp+12, bryan+13}.  This is mainly due to a change
in both the type and shape of orbits \citep{valluri+10}.  Nonetheless
models predict that haloes remain triaxial beyond $\sim 30-50$ kpc,
which, however, is a region that is poorly constrained by
observations.

The Sagittarius dwarf tidal stream, which extends to $\sim 60$ \kpc\
from the Galactic Centre, has been used to constrain the shape of the
Milky Way's halo with varying results.  Noting that the tidal debris
is distributed on a great circle, \citet{iba_etal_01} concluded that
the halo is nearly spherical.  Likewise \citet{fel_etal_06} argued
that the position of the bifurcation in the tidal stream, which they
interpreted as two wraps of the stream, can be explained if the halo
is close to spherical.  \citet{mar_etal_04} and \citet{joh_etal_05}
instead found a mildly oblate halo ($c/a \sim 0.9$) flattened in the
same sense as the disc.  \citet{helmi_04} meanwhile argued that the
trailing part of the Sagittarius Stream is too dynamically young to
provide a stringent constraint.  Using instead the leading stream,
\citet{helmi_04b} found evidence for a prolate halo with $c/a \sim
0.6$ and with its long axis perpendicular to the disc.  \citet{law+09}
were the first to demonstrate that simultaneously fitting the density
and kinematics of the Sagittarius Stream requires a triaxial (rather
than oblate/prolate) potential.  A surprising property of this
potential is that its intermediate axis is aligned with the short axis
of the disc (a relative orientation we will refer to as the
`intermediate-axis orientation').  \citet[][hereafter
LM10]{law_majewski10} presented a suite of $\sim 500$ $N$-body
simulations of the tidal disruption of the Sagittarius dwarf in a
fixed potential.  The simulations were compared with a large number of
constraints including (i) the position and velocity of the Sagittarius
dwarf with its velocity vector in the orbital plane of the young
trailing tail, (ii) the radial velocity and velocity dispersion in the
trailing tidal tail and (iii) the angular location, width and radial
velocities of the leading tail.  The best-fitting model is in the
intermediate-axis orientation with $(b/a)_\Phi = 0.99$ and $(c/a)_\Phi
= 0.72$ between 20 and 60 kpc, and with the major and minor axes in
the plane of the disc.  An analysis by \citet{deg_widrow13} that
varies also the parameters of the bulge$+$disc of the Milky Way still
finds a disc in the intermediate-axis orientation.  Recent extended
mapping of the Sagittarius Stream in the Southern Galactic hemisphere
finds a stream consistent with the LM10 model \citep{slater+13}.

Triaxial potentials are populated by box orbits (which get arbitrarily
close to the centre of the potential) and tube orbits (which have a
fixed sense of rotation relative to one of the principal axes).  The
stability of tube orbits about each of the principal axes of a
triaxial potential has been studied extensively: tube orbits are
stable around the short and long axes, but not around the intermediate
axis \citep{hei_sch_79, goo_sch_81, wil_jam_82}, even when planar
\citep{ada_etal_07, carpintero_muzzio11}.  Figure rotation gives rise
to warped planes of stable loop orbits capable of supporting discs
\citep{binney_78, heisler+82, magnenat_82, lake_norman_83, durisen+83,
  steiman-cameron_durisen_84, martinet_dezeeuw_88, habe_ikeuchi_85,
  habe_ikeuchi_88}, but the level of figure rotation of dark matter
haloes is found to be on average $\la10\degrees\Gyr^{-1}$ in
cosmological dark-matter-only simulations \citep{bailin_steinmetz_04}.
In semi-cosmological models, \citet{aumer_white13} showed that discs
are most stable when the angular momentum is aligned with the minor
axis of the halo.  The model of LM10 therefore challenges the view
that the instability of intermediate-axis tube (IAT) orbits prohibits
discs from forming in this orientation.  One way in which this
discrepancy might be resolved is if in the vicinity of the disc it
dominates the net potential, which becomes flattened like the disc.
Then the near-circular orbits in the disc are orbiting around the {\it
  local} short axis of the potential, and therefore in a stable
configuration (Johnston, private communication).


In this paper we show that discs are unable to persist in an
intermediate-axis orientation.  We use both simulations in which discs
are grown inside isolated triaxial haloes as well as a simulation of a
galaxy forming out of gas with angular momentum about the intermediate
axis of a triaxial halo.  In Section \ref{sec:numerics} we discuss the
methods used in this paper, including the initial conditions of the
stars, dark matter and gas.  Section \ref{sec:results} presents the
evolution of the models.  We draw our conclusions in Section
\ref{sec:conc}.  Appendix \ref{haloorbits} presents our interpretation
for why the intermediate-axis orientation is unstable based on an
orbital study.


\section{Numerical Methods}
\label{sec:numerics}

\subsection{Constructing collisionless initial conditions}

As in \citet{debattista+08}, we formed triaxial haloes via the merger
of three or more spherical haloes \citep{moo_etal_04}.  The mergers,
and all subsequent collisionless simulations, were evolved with {\sc
  pkdgrav} \citep{stadel_phd}, an efficient, multi-stepping, parallel
treecode.  The spherical haloes were generated from a distribution
function using the method of \citet{kmm04} with each halo composed of
two mass species arranged on shells.  The outer shell has more massive
particles than the inner one, increasing the effective resolution in
the centre.  As shown in \citet{debattista+08}, a large part of the
particle mass segregation persists after the mergers and the inner
region remains dominated by the lower mass particles.

\begin{figure}
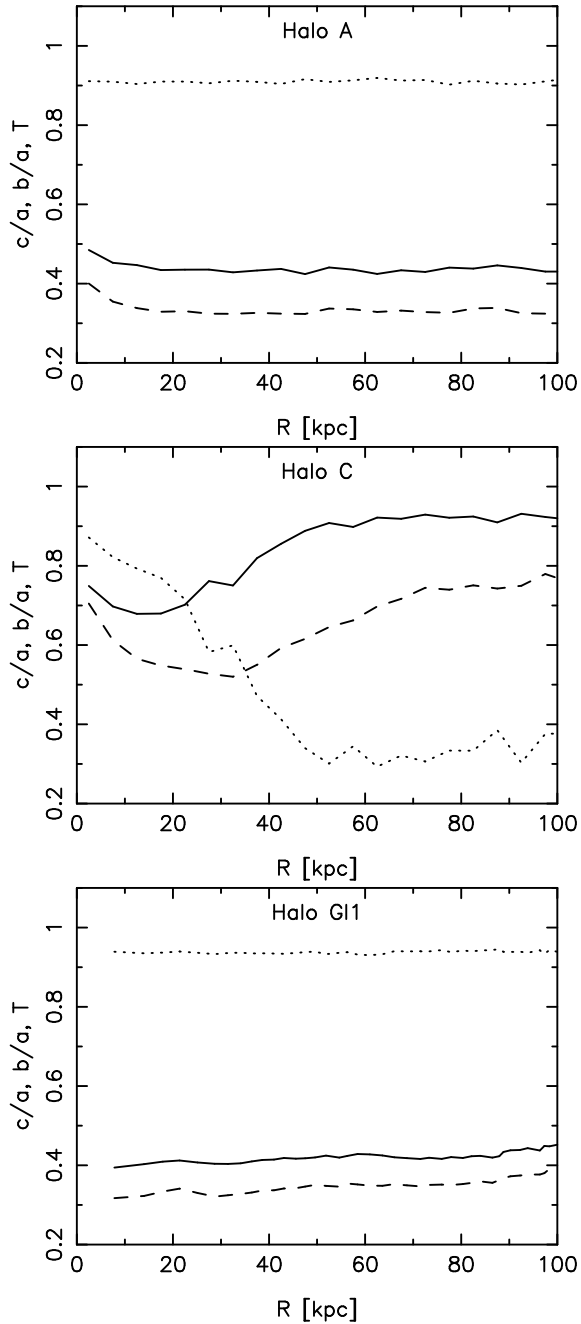

\includegraphics[angle=-90.,width=0.9\hsize]{figs/haloshapesA.ps}
\includegraphics[angle=-90.,width=0.9\hsize]{figs/haloshapesC.ps}
\includegraphics[angle=-90.,width=0.9\hsize]{figs/haloshapeGT.ps}
\caption{Density shape of dark matter haloes A (top), C (middle)
  and GI1 (bottom) before any of the discs/star formation are
  introduced.  Solid, dashed and dotted lines show $b/a$, $c/a$ and
  $T$, respectively.
  \label{fig:haloeshapes}}
\end{figure}

We produced two dark-matter-only triaxial haloes, which we refer to as
A and C; halo A was presented already in \citet{debattista+08}.  These
haloes were constructed from two consecutive mergers.  In both cases
the first merger placed two identical spherical concentration ${\cal
  C}=10$ haloes 800 kpc apart approaching each other at 50 \kms,
producing a prolate merged halo.  Halo A was generated by the head-on
merger of two copies of this remnant halo starting at rest 400 kpc
apart.  For halo C, after the first merger, a third spherical halo,
with ${\cal C}=20$, was merged from 100 kpc along the first remnant's
minor axis.  This ${\cal C}=20$ halo itself had two mass species
different from those of the ${\cal C}=10$ halo.  The top two panels of
Fig. \ref{fig:haloeshapes} plot the shape and triaxiality of these
two haloes, measured as described in \citet{debattista+08} \citep[see
also][]{zemp+11}, before any discs are introduced.  The triaxiality
parameter is defined as $T = (a^2 - b^2)/(a^2 - c^2)$
\citep{fra_etal_91}.  Halo A is highly prolate but has only a mild
triaxiality $T \sim 0.9$; its shape however is very constant out to
100 kpc.  Halo C instead has a radially varying $T$ ranging from $\sim
0.9$ at the centre to $\sim 0.3$ at 50 kpc.  Halo C is considerably
rounder than halo A everywhere within the inner 100 kpc.  Table
\ref{tab:haloes} lists the properties of the haloes\footnote{We use a
  different convention from \citet{debattista+08} and
  \citet{valluri+10, valluri+12}, who used the radius at which $\rho =
  200 \rho_{crit}$.  Here $r_{\rm 200}$ is the radius within which the
  enclosed mass has average density $200 \rho_{crit}$.}.

The outer particles are $\sim 19\times$ more massive than the inner
ones in halo A.  Halo C has 2 additional mass species which came with
the ${\cal C}=20$ halo: $\sim 1.8\times$ and $\sim 16\times$ more
massive than the low-mass particles in the ${\cal C}=10$ halo.  Both
the initial spherical halo with ${\cal C}=10$ and the one with ${\cal
  C}=20$ each had one million particles, equally divided between the
two mass species.  Thus halo A has four million particles while halo C
has three million.  We used a softening parameter $\epsilon = 0.1\kpc$
($\epsilon = 0.5\kpc$) for low (high) mass particles in both the
${\cal C}=10$ and the ${\cal C}=20$ spherical haloes.

\begin{centering}
\begin{table}
\vbox{\hfil
\begin{tabular}{cccccccc}\hline 
\multicolumn{1}{c}{Halo} &
\multicolumn{1}{c}{$N_{\rm p}$} &
\multicolumn{1}{c}{$N_{\rm g}$} &
\multicolumn{1}{c}{$M_{\rm 200}$} &
\multicolumn{1}{c}{$r_{\rm 200}$} &
\multicolumn{1}{c}{$b/a$} &
\multicolumn{1}{c}{$c/a$} \\

 & $(10^6)$ & $(10^6)$ & ($10^{12} \msun$) & (kpc) & & \\ \hline

A   & 3.3 & - & 6.3 & 379 & 0.45 & 0.35 \\ 
C   & 2.6 & - & 5.1 & 355 & 0.7 & 0.6 \\ 
GI1 & 2.8 & 2.7 & 3.2 & 304 & 0.4 & 0.32 \\ \hline 
\end{tabular}
\hfil}
\caption{
  The haloes used in the simulations.  The properties listed are for the
  halo after the last merger and before the discs have been grown.
  $N_{\rm p}$ and $N_{\rm g}$ are the number of dark matter and gas particles within
  $r_{\rm 200}$, and $M_{\rm 200}$ is the halo mass within the virial
  radius, $r_{\rm 200}$.  Density axes ratios $b/a$ and $c/a$ are by-eye
  averaged over the inner 20 kpc (see Fig. \ref{fig:haloeshapes}).}
\label{tab:haloes}
\end{table}
\end{centering}

Once we produced the triaxial haloes, we inserted a disc of particles
which initially remained rigid.  The disc distribution was, in all
cases, exponential with scale-length $R_{\rm d} = 3 \kpc$ and Gaussian
scale-height $z_{\rm d} = 0.05 R_{\rm d}$.  The discs are composed of
$300 000$ equal-mass particles.  Initially the disc has negligible mass
but this grows linearly over 5 Gyr.  During this time, the halo
particles are free to move and achieve equilibrium with the growing
disc.

The disc in halo A is grown to a mass of $1.75 \times 10^{11} \msun$.
The disc is placed in an intermediate-axis orientation and we
therefore refer to this model as model IA1.  \citet{valluri+12}
presented an orbital analysis of the halo in this model at $t=0$;
there the model is also referred to as IA1.  For some of our analysis,
we also present a version of this model with the disc at a mass of
only $7 \times 10^{10} \msun$, which we refer to as model IA2.
The disc in halo C instead is placed with its short axis along the
halo's long axis, so we refer to it as model LC1.  This disc has a
final mass of $1.4 \times 10^{11} \msun$.  The high disc masses in
models IA1 and LC1 allow us argue that even a high mass does not offer
a disc protection against the unstable intermediate-axis orientation,
but we have checked that lower mass discs (including in IA2) are also
unstable in this orientation.  
Disc particles in model IA1 each has a softening $\epsilon = 100\pc$,
while disc particles in models IA2 and LC1 have $\epsilon = 60\pc$.

We set the kinematics of the final discs to give constant $z_{\rm d}$
and Toomre $Q = 1.5$, as described in \citet{deb_sel_00}.  For this we
calculate the potential using a hybrid polar-grid code with the disc
on a cylindrical grid and the halo on a spherical grid
\citep{sellwo_03}.  In setting the disc kinematics, we azimuthally
average radial and vertical forces; thus our discs are initially not
in perfect equilibrium.  Equilibrium is quickly established once the
disc particles are free to move.  In these simulations $t=0$
corresponds to the time at which we set the disc kinematics.  {\sc
  pkdgrav} is a multi-stepping tree code, with time-steps refined such
that $\delta t = \Delta t/2^n < \eta (\epsilon/a_{\rm g})^{1/2}$,
where $\epsilon$ is the softening and $a_{\rm g}$ is the acceleration
at a particle's current position.  We use base time-step $\Delta t =
5$ Myr, $\eta = 0.2$ and set the opening angle of the treecode to
$\theta = 0.7$ in all cases.

\subsection{Initial conditions with gas}
\label{ssec:gasics}

We also present a simulation of a disc forming out of gas rotating
about the intermediate axis of a triaxial halo, which we refer to as
model GI1.  As did \citet{aumer_white13}, in our initial experiments
we found that arbitrarily inserting rotating gas haloes within
pre-existing non-spherical dark matter haloes leads to a substantial
loss of gas angular momentum.  Our approach therefore is to include
the gas, which is not allowed to cool, right from the start while
merging haloes to produce the triaxial system.  We first set up a
prolate halo with an equilibrium gas distribution by merging two
spherical Navarro-Frenk-White (NFW) dark matter haloes as before.
Each of the spherical initial haloes has an embedded spherical hot gas
component containing $10\%$ of the total mass and following the same
density distribution.  The initial haloes have been described in
\citet{rok_08a}: each dark matter halo has a mass within the virial
radius of $10^{12} \msun$.  A temperature gradient in each halo
ensures an initial gas pressure equilibrium for an adiabatic equation
of state.  Gas velocities are initialized to give a spin parameter of
$\lambda = 0.039$ \citep{bul_etal_angmom_01, mac_etal_07b}, with
specific angular momentum $j \propto R$, where $R$ is the cylindrical
radius.  Each halo used $10^6$ particles in each of the gas and dark
components.  Gas particles initially have masses $1.4\times 10^5
\msun$ and softening 50 pc, the latter inherited by the star
particles, while dark matter particles come in two mass flavours
($10^6\msun$ and $3.5\times10^6 \msun$ inside and outside 200 kpc,
respectively) and with a softening of 100 pc.  The two haloes are
placed 500 kpc apart along the $x$-axis and are initially moving
towards each other at a relative velocity of 100 \kms.

\begin{figure}
\includegraphics[angle=0.,width=\hsize]{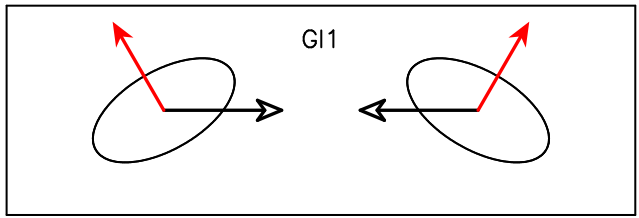}
\caption{ Cartoon representation of the merger geometry that produces
  a triaxial halo with gas angular momentum along the intermediate
  axis, model GI1.  The (red) filled arrows indicate the orientation
  of the gas angular momentum relative to the two merging prolate
  haloes while the (black) open arrows indicate the relative motion of
  the two haloes.}
\label{fig:mergergeometry}
\end{figure}

After the first merger the resulting halo is prolate, elongated along
the $x$-axis, with $\left<c/a\right> \simeq 0.65$ and angular momentum
along the short ($z$) axis.  We produce a triaxial halo by merging two
copies of this prolate system (for a total of $4\times 10^6$ particles
in each of the gas and dark matter components).  In order to align the
gas angular momentum with the intermediate axis of the halo we first
rotate the prolate system about the long axis so the angular momentum
vector is along the $y$-axis, then rotate two copies of the prolate
halo about the $z$-axis by $+30\degrees$ in one case and by
$-30\degrees$ in the other.  This merger geometry for the two prolate
haloes is illustrated in Fig. \ref{fig:mergergeometry}.  Merging
these two haloes from a separation of 500 kpc along the $x$-axis with a
relative velocity of 100 \kms\ produces a quite prolate halo with only
a very mild triaxiality ($T \sim 0.93$ within the inner 100 kpc), as
shown in the bottom panel of Fig. \ref{fig:haloeshapes}.
 
This simulation was evolved with {\sc gasoline} \citep{gasoline}, the
smooth particle hydrodynamics (SPH) version of {\sc pkdgrav}.  We use
a base time-step of 10 Myr with a refinement parameter $\eta = 0.175$.
During the mergers, and for some time after, we evolve the gas
adiabatically without cooling or star formation.  After, we switch on
gas cooling, star formation and stellar feedback using the
prescriptions of \citet{sti_etal_06}.
A gas particle undergoes star formation if it has number density $n >
0.1 {\mathrm cm}^{-3}$, temperature $T < 15,000$ K and is part of a
converging flow; efficiency of star formation is 0.05, \ie\ 5\% of gas
particles eligible to form stars do so per dynamical time.  Star
particles form with an initial mass of $1/3$ that of the parent gas
particle, which at our resolution corresponds to $4.6 \times 10^4
\msun$.  Gas particles can spawn multiple star particles but once they
drop below $1/5$ of their initial mass the remaining mass is
distributed amongst the nearest neighbours, leading to a decreasing
number of gas particles.  Each star particle represents an entire
stellar population with a Miller-Scalo \citep{miller_scalo79} initial
mass function. The evolution of star particles includes asymptotic
gian brach (AGB) stellar winds and feedback from Type II and Type Ia
supernovae, with their energy injected into the interstellar medium
(ISM). Each supernova releases $4 \times 10^{50}$ erg into the ISM.
The effect of the supernovae explosions is modelled as a subgrid
prescription for a blast wave propagating through the ISM
\citep{sti_etal_06}.  We again use an opening angle of $\theta = 0.7$.
The timestep of gas particles also satisfies the condition $\delta
t_{gas} = \eta_{courant} h/[(1+\alpha)c + \beta\mu_{max}]$, where
$\eta_{courant} = 0.4$, $h$ is the SPH smoothing length, $\alpha$ is
the shear coefficient, which is set to 1, $\beta=2$ is the viscosity
coefficient and $\mu_{max}$ is described in \citet{gasoline}.  The SPH
kernel is defined using the 32 nearest neighbours.  Gas cooling is
calculated without taking into account the gas metallicity.  These
prescriptions have been shown to lead to realistic Milky-Way-type
galaxies \citep{roskar+12, roskar+13}.  In this run, $t=0$ corresponds
to the time at which gas cooling is switched on and star formation
commences.

\subsection{Briggs figures}
\label{ssec:briggsfigs}

We use Briggs figures, originally introduced for studying warps
\citep{briggs_90}, to illustrate disc tilting in the simulations.  A
Briggs figure is a 2D polar coordinate representation of the direction
of vectors.  We decompose the stellar discs into five concentric rings
of equal width extending to a radius of 15 \kpc\ and for each ring
plot the direction of the angular momentum vector in 2D cylindrical
polar coordinates.  The tilt of the angular momentum vector from some
fiducial $z$-axis, $\theta$, is plotted as the radial coordinate,
while the angle from some fiducial $x$-axis, $\phi$, is plotted as the
angle coordinate.  Briggs figures are useful for showing the evolution
of disc orientation provided that the axes with respect to which the
angles $\theta$ and $\phi$ are defined are kept fixed.  Note that the
Briggs figure of a uniformly tilting disc consists of a set of
coincident points, indicating that the angular momentum of the disc is
everywhere aligned.  A differentially tilting (\ie\ warped) disc
instead is represented by non-coincident points.  In the collisionless
simulations we always set the $z$-axis to be the direction of the
angular momentum of the {\it initial} disc.  The reader is cautioned
that this is different from the convention adopted in
\citet{valluri+12}.


\section{Results}
\label{sec:results}

\subsection{Models IA1 and IA2}

\begin{figure}
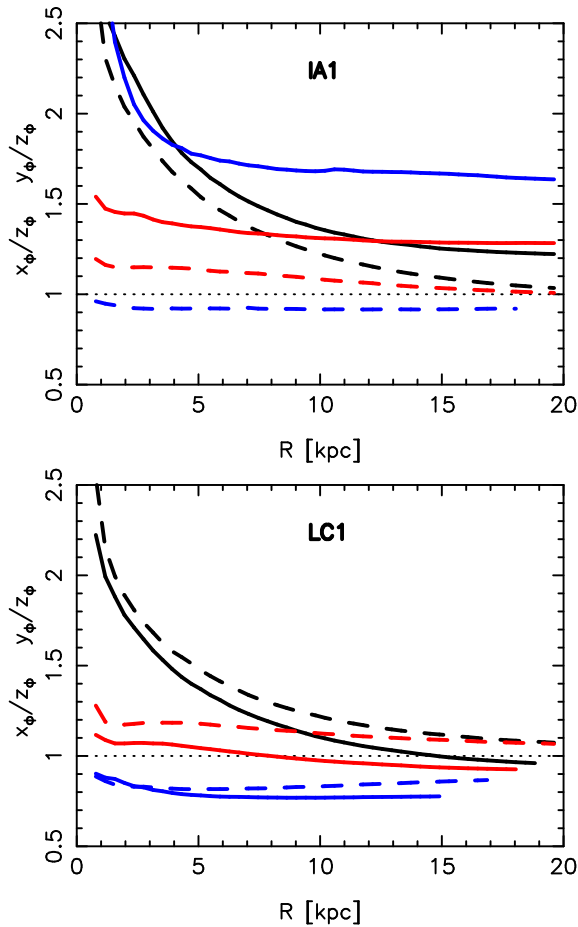

\includegraphics[angle=-90.,width=0.9\hsize]{figs/potshape534.ps}
\includegraphics[angle=-90.,width=0.9\hsize]{figs/potshape536.ps}
\caption{Equipotential axis ratios in models IA1 (top) and LC1 (bottom).
The solid lines show $x_\Phi/z_\Phi$ while the dashed lines show
$y_\Phi/z_\Phi$.  The thick blue lines correspond to the halo before
the disc is grown.  The black and red lines show the full and halo
potential shape after the disc is grown.  The dotted horizontal lines
indicate an axis ratio of unity.  The $z$-axis is perpendicular to the
disc at $t=0$.
\label{fig:potshapes}}
\end{figure}

In models IA1 and IA2 the $x$-axis is the pre-disc halo long axis,
while the $y$-axis is the short axis.  Once the disc is grown,
however, the inner halo is flattened to the extent that the disc's
vertical (\ie\ short) axis becomes the shortest axis of the inner
halo.  At larger radii the $x$ and $y$ axes continue to be the long
and short axes of the halo, so we use these to specify the axes
ordering.  We compute the potential in the $x=0$ and $y=0$ planes,
from which we measure the axes ratios of the potential by computing
the distance along each axis at which the potential takes particular
values.  The top panel of Fig. \ref{fig:potshapes} plots the
equipotential axis ratios $x_\Phi/z_\Phi$ and $y_\Phi/z_\Phi$ for IA1.
The pre-disc potential has $x_\Phi/z_\Phi > 1 > y_\Phi/z_\Phi
\Rightarrow x_\Phi > z_\Phi > y_\Phi$ but after the disc is grown,
within 20 kpc this becomes $x_\Phi > y_\Phi > z_\Phi$.  The mid-plane
potential has an ellipticity $\epsilon_\Phi \la 0.15$ within 20 kpc.

\begin{figure}
\centerline{\includegraphics[angle=0.,width=\hsize]{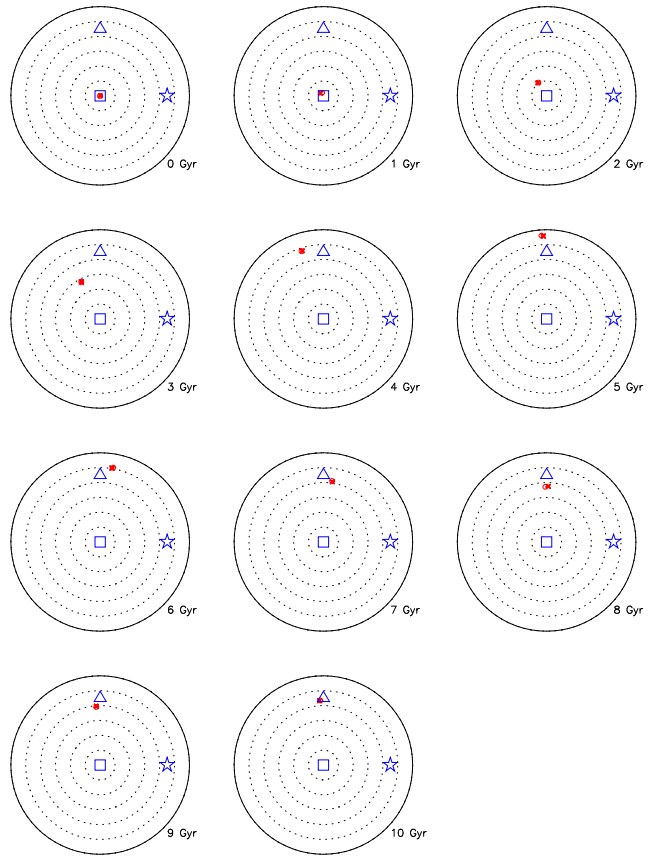}}
\caption{ Briggs figures (see Section \ref{ssec:briggsfigs} for an
  explanation of these figures) showing the evolution of run IA1 at 1
  \Gyr\ intervals.  Dotted circles are spaced at $20\degrees$
  intervals, with the outer solid circle corresponding to $\theta =
  120\degrees$.  The centre of the disc is indicated by the (red) open
  circle, while the remaining disc annuli are indicated by (red)
  crosses.  The open (blue) star, square and triangle symbols indicate
  the direction of the pre-disc halo long, intermediate and short
  axes, respectively.  
\label{fig:IA1briggs}}
\end{figure}

The evolution of run IA1 is shown in Fig. \ref{fig:IA1briggs}.  The
disc tilts by $90\degrees$ out of the initial plane within 4 \Gyr.
During this rapid tilting phase the disc does not warp substantially
or precess (which can be seen from the fact that the disc short axis
does not circulate about any axis).  At $t=4 \Gyr$ the disc has not
yet settled, having overshot the minor axis orientation to $\theta
\simeq 120\degrees$.  After 4 \Gyr\ the disc precesses about the short
axis while slowly settling into a short-axis orientation.  Throughout
this evolution, total angular momentum is conserved to better than
$1.5\%$, with angular momentum exchanged between the disc and the
halo.  Other than the disc tilting more rapidly, the lower disc mass
run IA2 evolves similar to run IA1.

IAT orbits are unstable \citep[e.g.][p.  263]{bin_tre_08}.  If the
disc is perpendicular to the intermediate axis of the potential, then
its stars would be on IAT orbits, which would render them unstable.
As Fig. \ref{fig:potshapes} shows, after the disc has grown the net
potential becomes so vertically flattened that the $z$-axis becomes
the shortest axis of the potential in the disc's vicinity.  This is
the case also if just the halo potential is considered.  Therefore the
disc tube orbits are stable because they are cocooned inside a
vertically flattened halo and circulate about the shortest axes of
their local potential.  We confirm this by repeating the simulation
with the halo particles frozen in place in model IA2.  Then the disc
does not tilt during 5 Gyr.

\begin{figure}
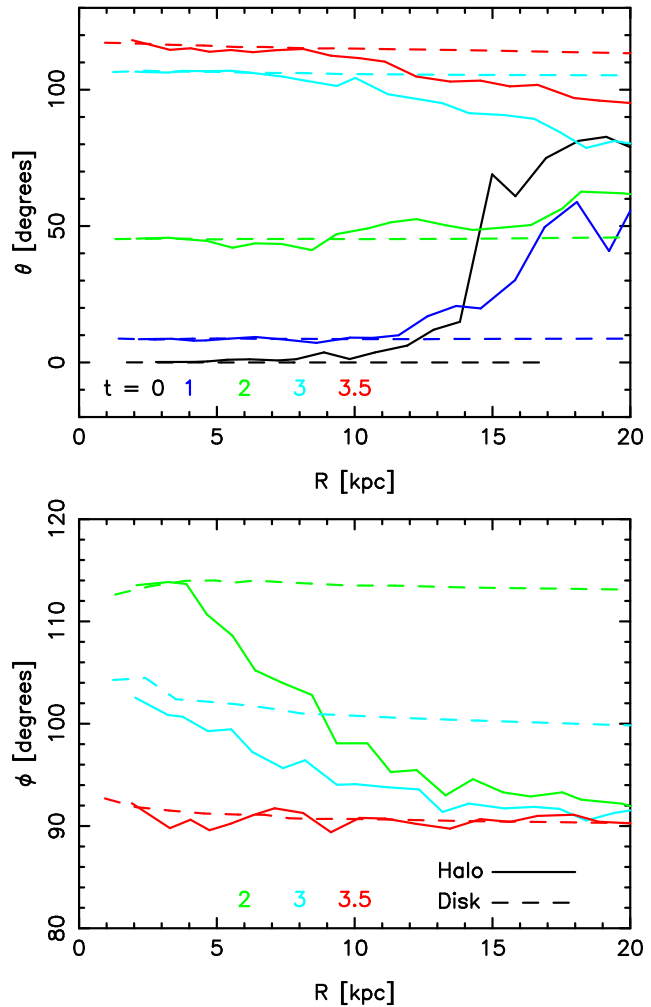

\centerline{\includegraphics[angle=-90.,width=\hsize]{figs/run534Adiskvshalo1.ps}}
\centerline{\includegraphics[angle=-90.,width=\hsize]{figs/run534Adiskvshalo2.ps}}
\caption{
Evolution of the relative orientation of the disc and inner halo in
model IA2.  The disc tilts rapidly from its initial orientation.  The
solid and dashed lines show the halo and disc orientations,
respectively, at different times, as indicated at bottom.  The black,
blue, green, cyan and red lines indicate $t=0$, 1, 2, 3 and 3.5 Gyr,
respectively. The $z$-axis relative to which $\theta$ is measured is
perpendicular to the {\it initial} disc, while the $x$-axis, which
defines $\phi = 0\degrees$, is the long axis.
\label{fig:relativetilt}}
\end{figure}

The instability must therefore reside in the halo.  In Appendix
\ref{haloorbits} we present evidence that the instability is driven by
the response of tube orbits to a potential with a radially varying
orientation.  Because the halo has negligible angular momentum, it
tilts without precessing, shepherding the disc along with it.
Evidence that the halo is driving the tilting of the disc comes also
from the small angular displacement between the disc and the inner
halo.  Close examination of Fig. \ref{fig:IA1briggs} shows that during
the tilting phase (2-4 Gyr), $\phi$ for the disc is not the same as
that for the halo minor axis.  In Fig. \ref{fig:IA1briggs}, the red
points mark the direction of the disc angular momentum; thus the disc
orientation during the tilting phase is ahead of (larger $\phi$) the
great circle between the intermediate and short axes, along which the
halo tilts.  In order to demonstrate this, we again use the lower disc
mass model IA2, since this distorts the inner halo to a lesser extent.
Fig. \ref{fig:relativetilt} shows the evolution of the direction of
the inner halo (solid lines) and of the disc (dashed lines) minor axes
separately, by plotting the tilt angle $\theta$ from the $z$-axis and
position angle $\phi$ from the $x$-axis.  The disc and halo tilt away
from the original vertical axis together, but the halo $\phi$ is
clearly closer to $\phi = 90\degrees$, corresponding to the outer halo
minor axis, than is the disc $\phi$.  Since the halo tilts almost
directly into the minor axis, the disc angle $\phi$ can be understood
as the disc misalignment relative to the halo needed to generate the
gravitational torque needed to reorient the disc.  Once the inner halo
has settled, the misalignment between the disc and the inner halo
leads to the damped precession seen after 4 Gyr.  Since the
instability is due to the halo, no matter how massive the disc becomes
(the halo-to-disc mass ratio within 15 kpc is 1.6), this orientation
can never be stable.

\subsection{Model LC1}

Before the disc is grown in run LC1, the direction vertical to the
disc is the long axis of halo C.  Fig. \ref{fig:potshapes} shows
that the ordering of the axes is $z_\Phi > y_\Phi > x_\Phi$ at this
stage, but once the disc is grown, the halo at $r \la 10$ kpc switches
orientation by $90\degrees$, so that the intermediate axis becomes the
axis orthogonal to the disc.  The combination of the disc and halo
potential then has $y_\Phi > x_\Phi > z_\Phi$ inside 15 kpc, and
$y_\Phi > z_\Phi > x_\Phi$ beyond.  Although the switch in the
principal axes of the density extends only to the inner halo, the flip
in the axes of the potential extends till at least 80 kpc.  The halo
flip is probably related to the accretion history of halo C which
included an accretion along the minor axis of a prolate halo.  Indeed
the inner halo major axis flips into the direction of the original
accretion event.  Thus while most of the disc is immersed
perpendicular to the short axis of the local potential, at larger
radii the disc short axis is along the intermediate axis of the
potential.  The resulting global potential has mid-plane potential
ellipticity $\epsilon_\Phi < 0.11$ everywhere within the inner 20 kpc.

\begin{figure}
\centerline{\includegraphics[angle=0.,width=\hsize]{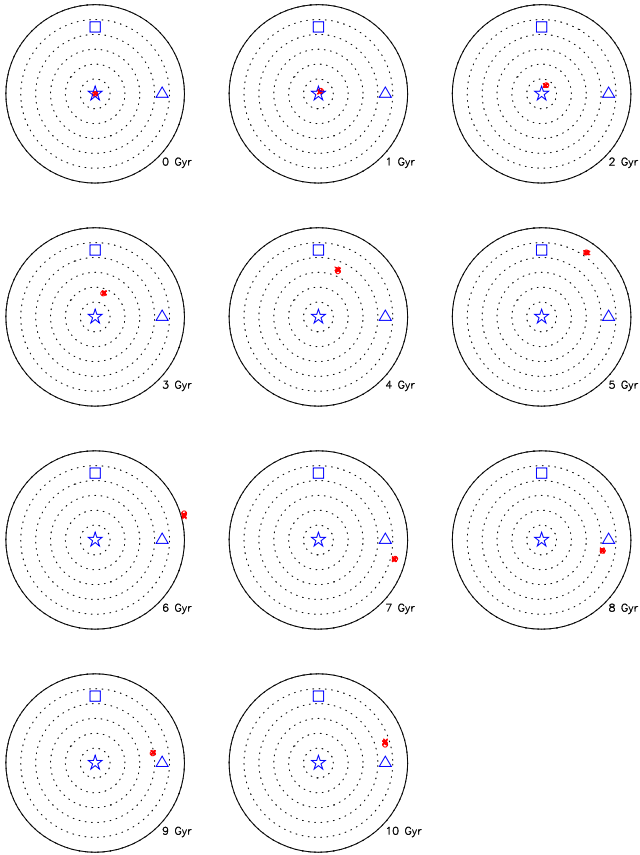}}
\caption{Briggs figures showing the evolution of run LC1 at 1 \Gyr\ intervals.
Dotted circles are spaced at $20\degrees$ intervals, with the outer
solid circle corresponding to $\theta = 120\degrees$.  The centre of
the disc is indicated by the (red) open circle, while the remaining
disc annuli are indicated by (red) crosses.  The open (blue) star,
square and triangle symbols indicate the direction of the pre-disc
halo long, intermediate and short axes, respectively.  In the inner
halo $y>z>x$ while in the outer halo $z>y>x$ once the disc is grown.
\label{fig:LC2briggs}}
\end{figure}

The disc in run LC1 tilts very rapidly, initially towards the original
intermediate-axis orientation and then dropping into a nearly
short-axis orientation, as shown in Fig. \ref{fig:LC2briggs}.  The
tilting rate reaches $\sim 30\degrees\Gyr^{-1}$ between 2 and 4 \Gyr.
This rapid, direct tilting is not accompanied by precession or
warping.  When we re-run the simulation with the halo frozen, the
outer disc still tilts and forms a polar ring, while the inner disc
tilts by only $\sim 15\degrees$.  Thus IAT orbits of stars in the
outer disc region are highly unstable.  However the entire disc is not
tilting because of this instability.  Given the lack of precession
when the disc is live, we conclude that the inner halo of run LC1 is
also in an unstable orientation, much as in run IA1.

\begin{figure}
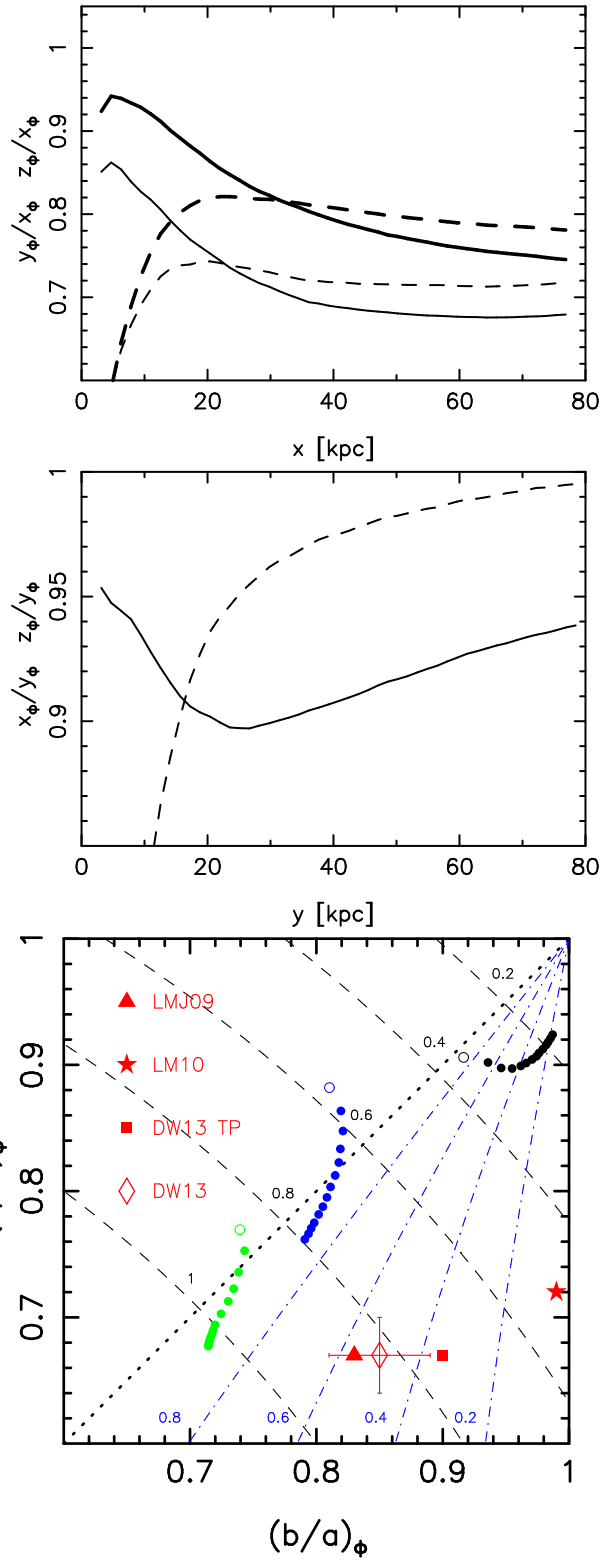

\includegraphics[angle=-90.,width=\hsize]{figs/compwithmods_a.ps}
\includegraphics[angle=-90.,width=\hsize]{figs/compwithmods_b.ps}
\includegraphics[angle=-90.,width=1.35\hsize]{figs/compwithmods_c.ps}
\caption{Top: profile of axes ratios of the potential at $t=0$ in
  models IA1 (thick lines) and IA2 (thin lines).  The solid (dashed)
  line shows $y_\Phi/x_\Phi$ ($z_\Phi/x_\Phi$).  Middle: profile of
  axes ratios of the potential at $t=0$ in model LC1.  The solid
  (dashed) line shows $x_\Phi/y_\Phi$ ($z_\Phi/y_\Phi$).  Bottom:
  potential axes ratios of LC1 (black circles), IA1 (blue circles) and
  IA2 (green circles) in the radial range $16 \leq r/\mathrm{kpc}
  \leq 60$ (with the open blue circle showing $16$ kpc and the filled
  circles showing larger radii) and Milky Way models.  Dashed lines
  are contours of deviations from sphericity, $\xi$, as defined in the
  text, while dot-dashed (blue) lines show contours of constant $T$.}
\label{fig:536Cpotshape}
\end{figure}

The middle panel of Fig. \ref{fig:536Cpotshape} shows the radial
profile of the potential axis ratios, $x_\Phi/y_\Phi$ and
$z_\Phi/y_\Phi$ to 80 kpc.  The longest axis of the potential
is the $y$-axis (recall that the axis vertical to the disc is $z$).
Beyond $\sim 15$ kpc, the potential intermediate axis is the $z$-axis
(\ie\ perpendicular to the initial disc) and its shape, while not
constant, does not vary substantially with radius.

\subsection{Comparison with previous models}

The bottom panel of Fig. \ref{fig:536Cpotshape} compares the shapes of
models IA1, IA2 and LC1 with the Milky Way potential in the
\citet{law+09}, LM10 and \citet{deg_widrow13} models.  Model LC1 has
larger (\ie\ rounder) $(c/a)_\Phi$ than all these models, while
$(b/a)_\Phi$ is comparable to the best LM10 and \citet{deg_widrow13}
TP models.  For a spherical potential, $(b/a)_\Phi^2 + (c/a)_\Phi^2 =
2$; we measure deviation from sphericity as $\xi = 2 - (b/a)_\Phi^2 -
(c/a)_\Phi^2$.  The bottom panel of Fig. \ref{fig:536Cpotshape} plots
contours of $\xi$ which clearly shows that the potential in LC1 is
more nearly spherical in this region than are the Milky Way models.
The instability of model LC1 is therefore very probably shared by all
these Milky Way models.

\subsection{Model GI1}
\label{ssec:gasrun}

\begin{figure}
\centerline{\includegraphics[angle=0.,width=\hsize]{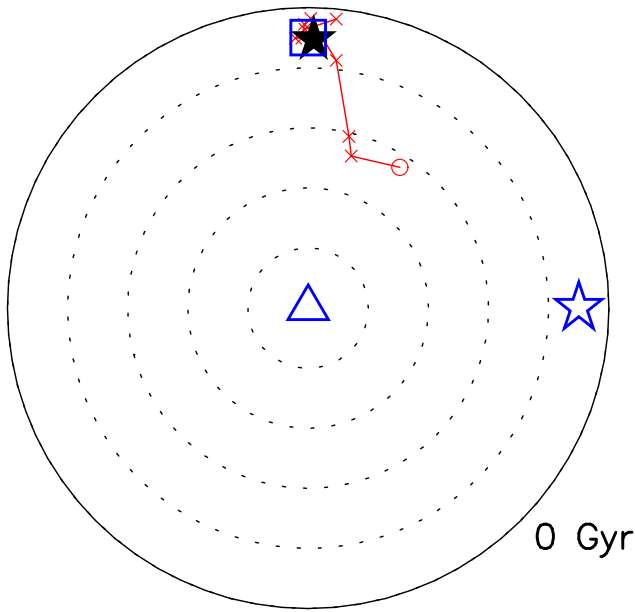}}
\caption{
Briggs figure for the gas within the inner 100 kpc in model GI1 at
$t=0$, before any star formation.  Dotted circles are spaced at
$20\degrees$ intervals, with the outer solid circle corresponding to
$\theta = 100\degrees$.  The centre of the gas halo is indicated by
the (red) open circle, while the remaining shells are indicated by
(red) crosses.  Each shell is 10 kpc wide.  The open (blue) star,
square and triangle symbols indicate the direction of the pre-disc
halo long, intermediate and short axes, respectively.  The (black) filled
star represents the orientation of the total gas angular
momentum within this volume.
\label{fig:run710gasbriggs}}
\end{figure}

Fig. \ref{fig:run710gasbriggs} shows the initial angular momentum of
the gas within the inner 100 kpc of model GI1.  The total angular
momentum within this region is very well aligned with the intermediate
axis of the halo.  Only within 30 kpc is the gas angular momentum not
in this orientation, but this corresponds to a tiny fraction of the
total angular momentum of this gas.

\begin{figure}
\centerline{\includegraphics[angle=0.,width=\hsize]{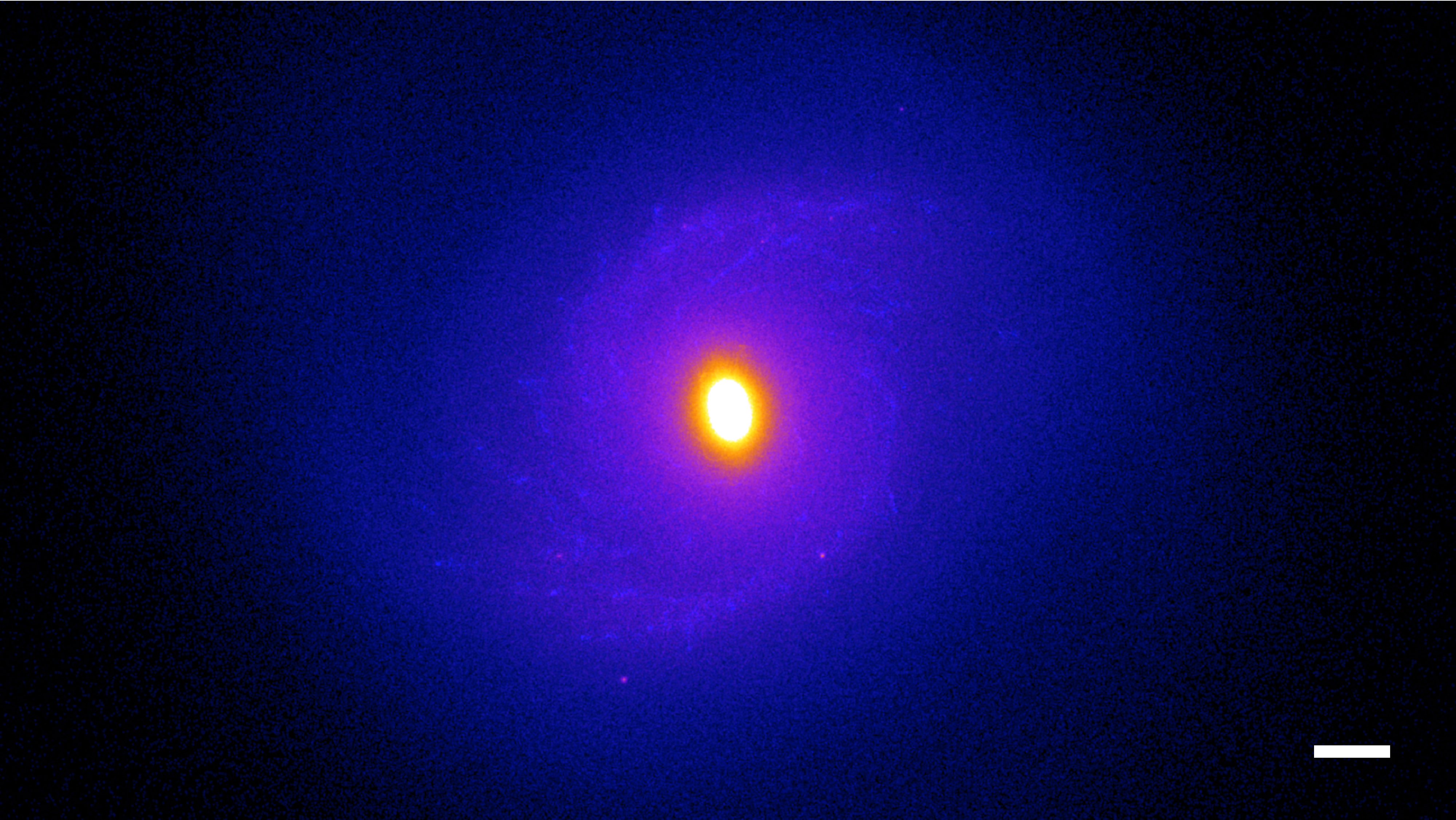}}
\centerline{\includegraphics[angle=0.,width=\hsize]{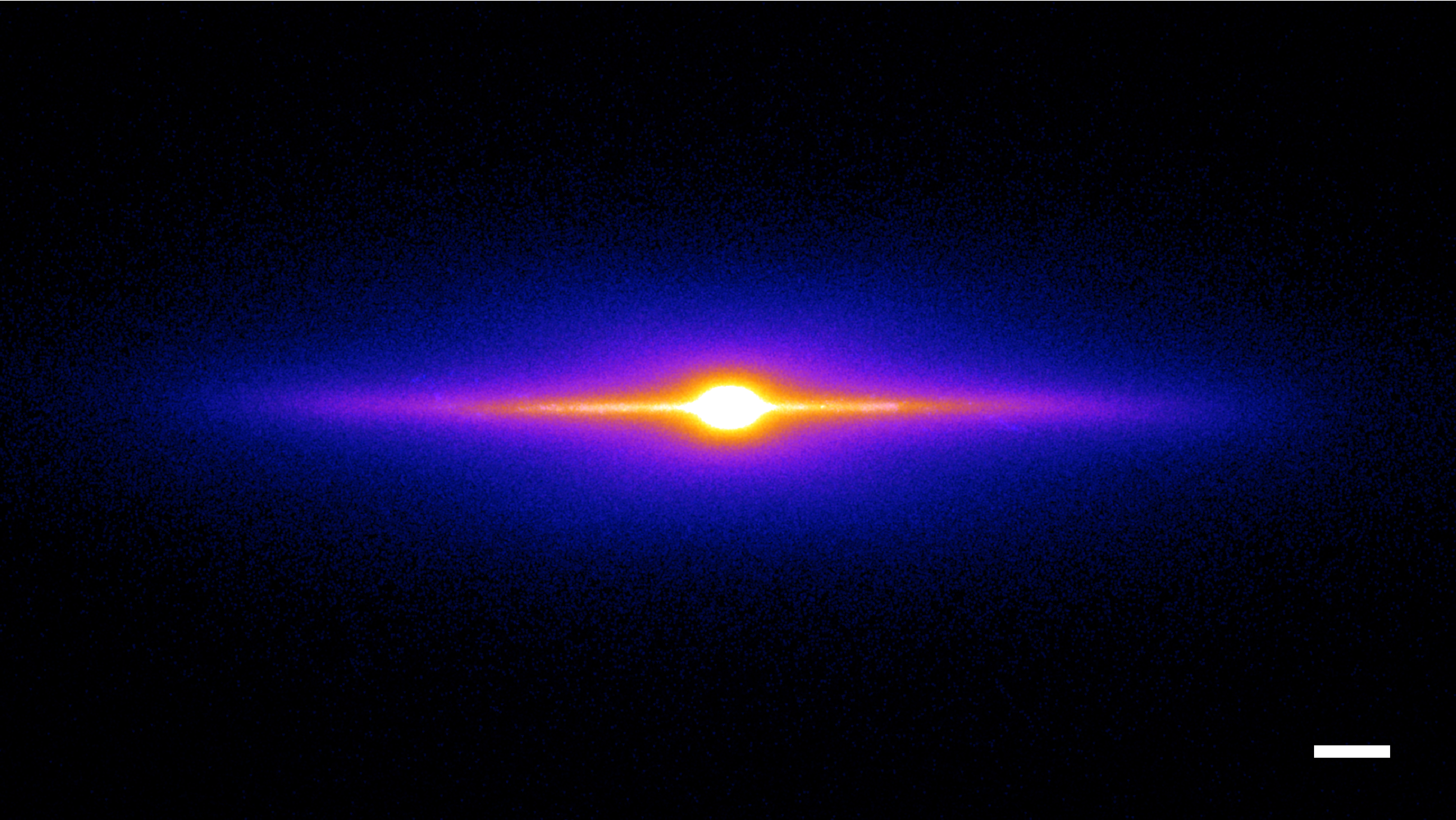}}
\caption{The stellar$+$gas disc of run GI1 at 6 Gyr as seen
    face-on (top) and edge-on (bottom).  The scale bar in the bottom right-hand
    corner indicates 1 kpc.
    \label{fig:run710disc}}
\end{figure}

\begin{figure}
\centerline{\includegraphics[angle=-90.,width=\hsize]{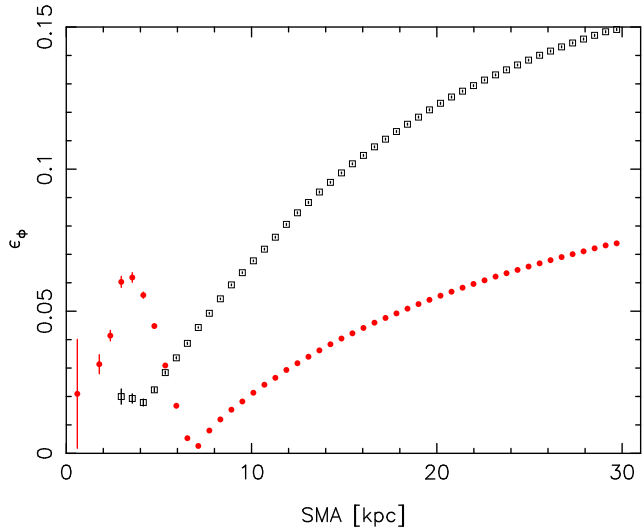}}
\caption{Ellipticity of the potential in the disc mid-plane,
    $\epsilon_\Phi$, for model GI1, plotted as a function of the
    semimajor axis.  The (black) open squares are at $2.5 \Gyr$, while
    the (red) filled circles show $6 \Gyr$.
    \label{fig:run710potell}}
\end{figure}

During the first 2 Gyr of evolution after gas cooling and star
formation are turned on the stellar disc is highly warped but by 2.5
Gyr it settles into a single plane.  Fig. \ref{fig:run710disc} shows
that by 6 Gyr a rapidly rotating thin stellar disc supporting spirals
has formed.  Fig. \ref{fig:run710potell} shows the profile of the
ellipticity of the potential in the disc plane, $\epsilon_\Phi$,
measured using the task {\sc ELLIPSE} in {\sc IRAF}\footnote{{\sc
    IRAF} is distributed by National Optical Astronomy Observatory
  (NOAO), which is operated by AURA Inc., under contract with the
  National Science Foundation.}.  Out to 30 kpc $\epsilon_\Phi \la
0.15$ at 2.5 Gyr, when the disc first becomes coherent.  This
decreases to $\epsilon_\Phi \la 0.08$ by 6 Gyr.  Thus $\epsilon_\Phi$
satisfies the stringent constraint from the scatter in the
Tully-Fisher relation \citep{fra_dez_92}.  By 9 Gyr the stellar disc
reaches a mass of $\sim 2 \times 10^{11} \msun$.

\begin{figure}
\centerline{\includegraphics[angle=0.,width=\hsize]{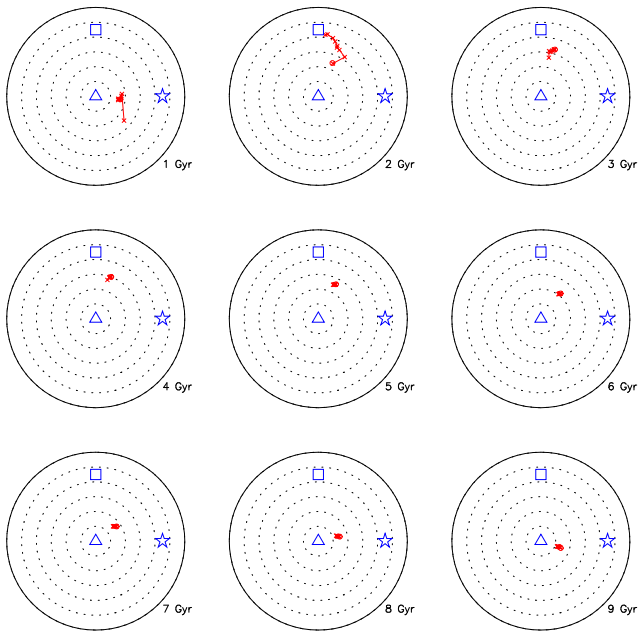}}
\caption{
Briggs figure for run GI1 at 1 \Gyr\ intervals.  Dotted circles are
spaced at $20\degrees$ intervals, with the outer solid circle
corresponding to $\theta = 120\degrees$.  The centre of the disc is
indicated by the (red) open circle, while the remaining disc annuli are
indicated by (red) crosses.  The open (blue) star, square and triangle
symbols indicate the direction of the pre-disc halo long, intermediate
and short axes, respectively.
\label{fig:run710briggs}}
\end{figure}

Fig. \ref{fig:run710briggs} shows the evolution of the disc
orientation.  The stellar disc never settles into an intermediate-axis
orientation; at 3 Gyr the disc is inclined by $\sim 30\degrees$ to
this axis, increasing to $\sim 100\degrees$ by 9 Gyr.  Thus even with
the global gas angular momentum aligned with the intermediate axis,
the disc cannot form in an intermediate-axis orientation even though
the halo is only very mildly triaxial, with $T \sim 0.93$ throughout
the inner 100 kpc before the disc forms.

\section{Discussion}
\label{sec:conc}

\begin{figure}
\includegraphics[angle=-90.,width=\hsize]{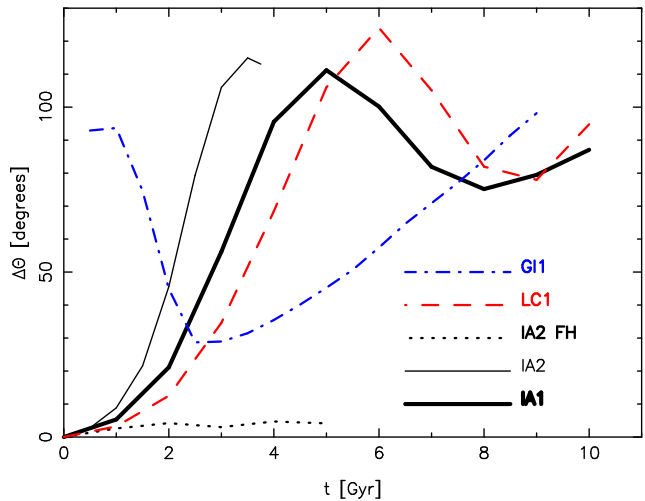}
\caption{ Tilting of the models away from the intermediate-axis
  orientation.  Different simulations are shown by different line
  styles as indicated.  Model IA2 FH corresponds to model IA2
    with halo particles frozen.  During the first 2 Gyr model GI1 is
  highly warped before it settles into a coherent plane.
  \label{fig:tilting}}
\end{figure}

We have shown that a disc can never remain with its minor axis aligned
with the intermediate axis of a triaxial halo (an `intermediate-axis
orientation').  This is shown in a different way in Fig.
\ref{fig:tilting}, which plots the evolution of the angle between the
stellar disc angular momentum and the halo's intermediate axis.  In
models IA1, IA2 and LC1 this angle increases rapidly until the disc is
nearly orthogonal.  In model GI1 the disc is initially chaotic, but
once it settles after 2.5 Gyr the angle increases throughout.  This
happens even if the disc cocoons itself by flattening the inner halo
such that the minor axis of the net potential is perpendicular to the
disc where it resides.  Such a vertically flattened inner halo is the
expected configuration within 20 kpc for the LM10 triaxial model of
the Milky Way (Johnston, private communication).  In that case, the
orbits of stars in the disc are stable.  However, a disc grown in an
intermediate-axis orientation gives rise to an instability {\it in the
  halo}.  As a result the inner halo tilts rapidly (within $\sim 4
\Gyr$), shepherding the disc along with it.  A hallmark of this
instability is that the disc tilts without precessing, as it stays
near equilibrium with the tilting inner halo.

We also showed, by means of a simulation with gas and star formation,
that even if the gas angular momentum is along the intermediate axis,
then the disc which forms is not in the intermediate-axis orientation.
This happens even if the halo is only very mildly triaxial: in model
GI1 the halo density has $T \simeq 0.93$.  We conclude that discs
cannot form in an intermediate-axis orientation, and even if they were
perturbed into such an orientation, they would not last long in it.
Since the instability resides in the halo, it also seems unlikely that
even more massive discs would be able inhibit it.

The shape of the LC1 potential is strongly varying inside $\sim 15$
kpc but this part of the potential is poorly constrained by the
Sagittarius Stream \citep[but see][for other constraints]{loebman+12}.
Beyond this radius, the potential shape varies quite slowly.  The
ratio $(b/a)_\Phi \ga 0.9$ which is not much different from the LM10
model, while $(c/a)_\Phi \sim 0.9$, which is larger than in the LM10
and \citet{deg_widrow13} models.  Thus the potential in model LC1 is
closer to spherical than the models of LM10 and \citet{deg_widrow13}.
The top and bottom panels of Fig.  \ref{fig:536Cpotshape} also show
the shape of the potential in models IA1 and IA2.  Both of these are
quite prolate, with model IA1 closer to spherical than the best Milky
Way model of DW13.  These less spherical Milky Way models would
therefore probably also be highly unstable.  Since the Milky Way has
not experienced strong interactions in the past few gigayears that
might have put it in an intermediate-axis orientation, it is very
unlikely to be in such an orientation.

Further difficulties for the Sagittarius Stream models come from their
failure to match the leading arm of the stream well, and to produce
the observed bifurcation \citep{belokurov+06}, which has now been
detected also in the trailing arm in the south \citep{koposov+12}.  We
note that the best-fitting model of LM10, while it does an excellent
job of fitting much of the observational data, still has
$\tilde{\chi}^2 = 3.4$ (but in comparison, their spherical halo has
$\tilde{\chi}^2 \simeq 9$).  In the past solutions of these problems
have been sought, unsuccessfully, in details of the Sagittarius dwarf
itself \citep[e.g.][]{penarrubia+10, penarrubia+11}.  Here we have
shown that triaxial models of the Milky Way which consistently find
the disc in an intermediate-axis orientation themselves can be ruled
out.

What then is the most promising way to improve Milky Way halo models
of the Sagittarius Stream?  The assumption of a constant shape within
the region of the Sagittarius Stream is unlikely to be correct;
however, halo shapes generally change sufficiently slowly beyond the
disc that this assumption amounts to measuring an average shape rather
than completely invalidating past models (note, for instance, how
small the variation in the shape of the potential of model LC1 is from
16 to 60 kpc in Fig. \ref{fig:536Cpotshape}).

\citet{ibata+13} showed that if the halo rotation curve is allowed to
increase to $\sim 300 \kms$ at 60 kpc that it is still possible to fit
the Sagittarius Stream by a spherical model.  This model still fails
to produce a bifurcation and results in a quite massive Milky Way
($2.6-3.1 \times 10^{12} \msun$).  As argued by \citet{ibata+13}, such
a model cannot be excluded by current observational constraints but it
would be unusual in $\Lambda$ cold dark matter ($\Lambda$CDM).
Nonetheless, more general density profiles are certainly highly
recommended for future models.

We propose here a different, and more natural, solution to the
problems of the Sagittarius Stream.  The models of
\citet{deg_widrow13} as well as those of \citet{law+09} vary the axes
ratios of the halo such that if the disc had been perpendicular to
either the short or the long axes of the halo then the models would
have been able to recover this; the fact that they did not means that
the Milky Way disc is not in either orientation.  We contend that the
assumption that the disc of the Milky Way is in one of the symmetry
planes of the halo {\it must} be incorrect.  The possibility that this
assumption can fail is clearly illustrated by our model GI1 which
shows that the disc does not need to be sitting in one of the
principal planes of a triaxial halo outside the region dominated by
the disc.  Indeed in cosmological simulations a decoupling between the
disc/inner halo and the outer halo is a common outcome
\citep{bailin+05, roskar+10}.  The most promising way to improve
future models of the Milky Way's halo shape from the Sagittarius
Stream is, therefore, the freedom for the disc to not be in one of the
symmetry planes of the halo.  Such models can be constrained further
by the cold tidal streams of lower mass progenitors, which can provide
more accurate tracers of the underlying potential
\citep{penarrubia+12, lux+12}.

While complicating efforts at understanding the halo, this orientation
nonetheless provides a unique opportunity to test the Modified
Newtonian Dynamics \citep[MOND;][]{mond, teves}.  If the Sagittarius
Stream requires a net potential that is tilted with respect to the
Milky Way disc, as we have argued, then this would constitute a
problem for MOND, which requires the short axis of the disc and of the
net potential to be parallel \citep[see also][]{buote_canizares94,
  read_moore05}.  The forthcoming generation of Milky Way surveys and
missions such as {\it Gaia} \citep{gaia} and the Large Synoptic Survey
Telescope \citep{ive_etal_08} will provide the data needed for much
more accurate modelling of the Milky Way's potential.

\bigskip
\noindent 
{\bf ACKNOWLEDGEMENTS}

\noindent

The collisionless simulations were performed at the Arctic Region
Supercomputing Center.  The simulations with gas were carried out at
the HPC facility of the University of Malta and at the HPC Facility at
the University of Central Lancashire.  VPD thanks the University of
Z\"urich for hospitality during part of this project.  Support for a
visit by Short Visit Grant \# 2442 within the framework of the ESF
Research Networking Programme entitled 'Computational Astrophysics and
Cosmology' is gratefully acknowledged.  We thank Kathryn Johnston for
discussion and for sharing unpublished results with us, and Nathan
Deg, Vasily Belokurov, Wyn Evans, Stacy McGaugh, Jorge Pe\~narrubia,
Justin Read, and Marcel Zemp for fruitful discussions.  We thank
Laurent Serge Noel for producing Fig. \ref{fig:run710disc}.  We thank
the anonymous referee for comments that helped improve this paper.
VPD thanks Bruno Debattista for the fun times that inspired the title
of this paper.  VPD is supported in part by STFC Consolidated grant \#
ST/J001341/1.  RR is supported by a Marie Curie Career Integration
Grant.  MV is supported by NSF grant AST-0908346 and by University of
Michigan's Elizabeth Crosby grant.

\bigskip 
\noindent

\bibliographystyle{aj}
\bibliography{allrefs}

\begin{thebibliography}{}

\bibitem[\protect\citeauthoryear{{Adams} et~al.}{{Adams}
  et~al.}{2007}]{ada_etal_07}
{Adams}, F.~C., {Bloch}, A.~M., {Butler}, S.~C., {Druce}, J.~M.,  \& {Ketchum},
  J.~A. 2007, \apj, 670, 1027

\bibitem[\protect\citeauthoryear{{Allgood} et~al.}{{Allgood}
  et~al.}{2006}]{all_etal_06}
{Allgood}, B., {Flores}, R.~A., {Primack}, J.~R., {Kravtsov}, A.~V.,
  {Wechsler}, R.~H., {Faltenbacher}, A.,  \& {Bullock}, J.~S. 2006, \mnras,
  367, 1781

\bibitem[\protect\citeauthoryear{{Andersen} et~al.}{{Andersen}
  et~al.}{2001}]{and_etal_01}
{Andersen}, D.~R., {Bershady}, M.~A., {Sparke}, L.~S., {Gallagher}, J.~S.,  \&
  {Wilcots}, E.~M. 2001, \apjl, 551, L131

\bibitem[\protect\citeauthoryear{{Aumer} \& {White}}{{Aumer} \&
  {White}}{2013}]{aumer_white13}
{Aumer}, M.,  \& {White}, S.~D.~M. 2013, \mnras, 428, 1055

\bibitem[\protect\citeauthoryear{{Bailin} et~al.}{{Bailin}
  et~al.}{2005}]{bailin+05}
{Bailin}, J., et~al. 2005, \apjl, 627, L17

\bibitem[\protect\citeauthoryear{{Bailin} \& {Steinmetz}}{{Bailin} \&
  {Steinmetz}}{2004}]{bailin_steinmetz_04}
{Bailin}, J.,  \& {Steinmetz}, M. 2004, \apj, 616, 27

\bibitem[\protect\citeauthoryear{{Bailin} \& {Steinmetz}}{{Bailin} \&
  {Steinmetz}}{2005}]{bai_ste_05}
{Bailin}, J.,  \& {Steinmetz}, M. 2005, \apj, 627, 647

\bibitem[\protect\citeauthoryear{{Banerjee} \& {Jog}}{{Banerjee} \&
  {Jog}}{2008}]{ban_jog_08}
{Banerjee}, A.,  \& {Jog}, C.~J. 2008, \apj, 685, 254

\bibitem[\protect\citeauthoryear{{Bardeen} et~al.}{{Bardeen}
  et~al.}{1986}]{bbks_86}
{Bardeen}, J.~M., {Bond}, J.~R., {Kaiser}, N.,  \& {Szalay}, A.~S. 1986, \apj,
  304, 15

\bibitem[\protect\citeauthoryear{{Barnes} \& {Sellwood}}{{Barnes} \&
  {Sellwood}}{2003}]{bar_sel_03}
{Barnes}, E.~I.,  \& {Sellwood}, J.~A. 2003, \aj, 125, 1164

\bibitem[\protect\citeauthoryear{{Barnes} \& {Efstathiou}}{{Barnes} \&
  {Efstathiou}}{1987}]{bar_efs_87}
{Barnes}, J.,  \& {Efstathiou}, G. 1987, \apj, 319, 575

\bibitem[\protect\citeauthoryear{{Bartelmann}, {Steinmetz}, \&
  {Weiss}}{{Bartelmann} et~al.}{1995}]{bar_etal_95}
{Bartelmann}, M., {Steinmetz}, M.,  \& {Weiss}, A. 1995, \aap, 297, 1

\bibitem[\protect\citeauthoryear{{Bekenstein}}{{Bekenstein}}{2004}]{teves}
{Bekenstein}, J.~D. 2004, \prd, 70, 083509

\bibitem[\protect\citeauthoryear{{Belokurov} et~al.}{{Belokurov}
  et~al.}{2006}]{belokurov+06}
{Belokurov}, V., et~al. 2006, \apjl, 642, L137

\bibitem[\protect\citeauthoryear{{Binney}}{{Binney}}{1978}]{binney_78}
{Binney}, J. 1978, \mnras, 183, 779

\bibitem[\protect\citeauthoryear{{Binney} \& {Tremaine}}{{Binney} \&
  {Tremaine}}{2008}]{bin_tre_08}
{Binney}, J.,  \& {Tremaine}, S. 2008, {Galactic Dynamics: Second Edition}
  (Galactic Dynamics: Second Edition, by James Binney and Scott Tremaine.~ISBN
  978-0-691-13026-2 (HB).~Published by Princeton University Press, Princeton,
  NJ USA, 2008.)

\bibitem[\protect\citeauthoryear{{Briggs}}{{Briggs}}{1990}]{briggs_90}
{Briggs}, F.~H. 1990, \apj, 352, 15

\bibitem[\protect\citeauthoryear{{Bryan} et~al.}{{Bryan}
  et~al.}{2013}]{bryan+13}
{Bryan}, S.~E., {Kay}, S.~T., {Duffy}, A.~R., {Schaye}, J., {Vecchia}, C.~D.,
  \& {Booth}, C.~M. 2013, \mnras, 429, 3316

\bibitem[\protect\citeauthoryear{{Bullock} et~al.}{{Bullock}
  et~al.}{2001}]{bul_etal_angmom_01}
{Bullock}, J.~S., {Dekel}, A., {Kolatt}, T.~S., {Kravtsov}, A.~V., {Klypin},
  A.~A., {Porciani}, C.,  \& {Primack}, J.~R. 2001, \apj, 555, 240

\bibitem[\protect\citeauthoryear{{Buote} \& {Canizares}}{{Buote} \&
  {Canizares}}{1994}]{buote_canizares94}
{Buote}, D.~A.,  \& {Canizares}, C.~R. 1994, \apj, 427, 86

\bibitem[\protect\citeauthoryear{{Buote} et~al.}{{Buote}
  et~al.}{2002}]{buo_etal_02}
{Buote}, D.~A., {Jeltema}, T.~E., {Canizares}, C.~R.,  \& {Garmire}, G.~P.
  2002, \apj, 577, 183

\bibitem[\protect\citeauthoryear{{Carpintero} \& {Aguilar}}{{Carpintero} \&
  {Aguilar}}{1998}]{carpintero_aguilar98}
{Carpintero}, D.~D.,  \& {Aguilar}, L.~A. 1998, \mnras, 298, 1

\bibitem[\protect\citeauthoryear{{Carpintero} \& {Muzzio}}{{Carpintero} \&
  {Muzzio}}{2012}]{carpintero_muzzio11}
{Carpintero}, D.~D.,  \& {Muzzio}, J.~C. 2012, Celestial Mechanics and
  Dynamical Astronomy, 112, 107

\bibitem[\protect\citeauthoryear{{Debattista}}{{Debattista}}{2003}]{debatt_03}
{Debattista}, V.~P. 2003, \mnras, 342, 1194

\bibitem[\protect\citeauthoryear{{Debattista} et~al.}{{Debattista}
  et~al.}{2008}]{debattista+08}
{Debattista}, V.~P., {Moore}, B., {Quinn}, T., {Kazantzidis}, S., {Maas}, R.,
  {Mayer}, L., {Read}, J.,  \& {Stadel}, J. 2008, \apj, 681, 1076

\bibitem[\protect\citeauthoryear{{Debattista} \& {Sellwood}}{{Debattista} \&
  {Sellwood}}{2000}]{deb_sel_00}
{Debattista}, V.~P.,  \& {Sellwood}, J.~A. 2000, \apj, 543, 704

\bibitem[\protect\citeauthoryear{{Deg} \& {Widrow}}{{Deg} \&
  {Widrow}}{2013}]{deg_widrow13}
{Deg}, N.,  \& {Widrow}, L. 2013, \mnras, 428, 912

\bibitem[\protect\citeauthoryear{{Diehl} \& {Statler}}{{Diehl} \&
  {Statler}}{2007}]{die_sta_07}
{Diehl}, S.,  \& {Statler}, T.~S. 2007, \apj, 668, 150

\bibitem[\protect\citeauthoryear{{Dubinski}}{{Dubinski}}{1994}]{dubins_94}
{Dubinski}, J. 1994, \apj, 431, 617

\bibitem[\protect\citeauthoryear{{Dubinski} \& {Carlberg}}{{Dubinski} \&
  {Carlberg}}{1991}]{dub_car_91}
{Dubinski}, J.,  \& {Carlberg}, R.~G. 1991, \apj, 378, 496

\bibitem[\protect\citeauthoryear{{Durisen} et~al.}{{Durisen}
  et~al.}{1983}]{durisen+83}
{Durisen}, R.~H., {Tohline}, J.~E., {Burns}, J.~A.,  \& {Dobrovolskis}, A.~R.
  1983, \apj, 264, 392

\bibitem[\protect\citeauthoryear{{Fellhauer} et~al.}{{Fellhauer}
  et~al.}{2006}]{fel_etal_06}
{Fellhauer}, M., et~al. 2006, \apj, 651, 167

\bibitem[\protect\citeauthoryear{{Franx} \& {de Zeeuw}}{{Franx} \& {de
  Zeeuw}}{1992}]{fra_dez_92}
{Franx}, M.,  \& {de Zeeuw}, T. 1992, \apjl, 392, L47

\bibitem[\protect\citeauthoryear{{Franx}, {Illingworth}, \& {de Zeeuw}}{{Franx}
  et~al.}{1991}]{fra_etal_91}
{Franx}, M., {Illingworth}, G.,  \& {de Zeeuw}, T. 1991, \apj, 383, 112

\bibitem[\protect\citeauthoryear{{Franx}, {van Gorkom}, \& {de Zeeuw}}{{Franx}
  et~al.}{1994}]{fra_etal_94}
{Franx}, M., {van Gorkom}, J.~H.,  \& {de Zeeuw}, T. 1994, \apj, 436, 642

\bibitem[\protect\citeauthoryear{{Frenk} et~al.}{{Frenk}
  et~al.}{1988}]{frenk_etal_88}
{Frenk}, C.~S., {White}, S.~D.~M., {Davis}, M.,  \& {Efstathiou}, G. 1988,
  \apj, 327, 507

\bibitem[\protect\citeauthoryear{{Goodman} \& {Schwarzschild}}{{Goodman} \&
  {Schwarzschild}}{1981}]{goo_sch_81}
{Goodman}, J.,  \& {Schwarzschild}, M. 1981, \apj, 245, 1087

\bibitem[\protect\citeauthoryear{{Habe} \& {Ikeuchi}}{{Habe} \&
  {Ikeuchi}}{1985}]{habe_ikeuchi_85}
{Habe}, A.,  \& {Ikeuchi}, S. 1985, \apj, 289, 540

\bibitem[\protect\citeauthoryear{{Habe} \& {Ikeuchi}}{{Habe} \&
  {Ikeuchi}}{1988}]{habe_ikeuchi_88}
{Habe}, A.,  \& {Ikeuchi}, S. 1988, \apj, 326, 84

\bibitem[\protect\citeauthoryear{{Heiligman} \& {Schwarzschild}}{{Heiligman} \&
  {Schwarzschild}}{1979}]{hei_sch_79}
{Heiligman}, G.,  \& {Schwarzschild}, M. 1979, \apj, 233, 872

\bibitem[\protect\citeauthoryear{{Heisler}, {Merritt}, \&
  {Schwarzschild}}{{Heisler} et~al.}{1982}]{heisler+82}
{Heisler}, J., {Merritt}, D.,  \& {Schwarzschild}, M. 1982, \apj, 258, 490

\bibitem[\protect\citeauthoryear{{Helmi}}{{Helmi}}{2004a}]{helmi_04}
{Helmi}, A. 2004a, \mnras, 351, 643

\bibitem[\protect\citeauthoryear{{Helmi}}{{Helmi}}{2004b}]{helmi_04b}
{Helmi}, A. 2004b, \apjl, 610, L97

\bibitem[\protect\citeauthoryear{{Huizinga} \& {van Albada}}{{Huizinga} \& {van
  Albada}}{1992}]{hui_van_92}
{Huizinga}, J.~E.,  \& {van Albada}, T.~S. 1992, \mnras, 254, 677

\bibitem[\protect\citeauthoryear{{Ibata} et~al.}{{Ibata}
  et~al.}{2001}]{iba_etal_01}
{Ibata}, R., {Lewis}, G.~F., {Irwin}, M., {Totten}, E.,  \& {Quinn}, T. 2001,
  \apj, 551, 294

\bibitem[\protect\citeauthoryear{{Ibata} et~al.}{{Ibata}
  et~al.}{2013}]{ibata+13}
{Ibata}, R., {Lewis}, G.~F., {Martin}, N.~F., {Bellazzini}, M.,  \& {Correnti},
  M. 2013, \apjl, 765, L15

\bibitem[\protect\citeauthoryear{{Iodice} et~al.}{{Iodice}
  et~al.}{2003}]{iod_etal_03}
{Iodice}, E., {Arnaboldi}, M., {Bournaud}, F., {Combes}, F., {Sparke}, L.~S.,
  {van Driel}, W.,  \& {Capaccioli}, M. 2003, \apj, 585, 730

\bibitem[\protect\citeauthoryear{{Ivezic} et~al.}{{Ivezic}
  et~al.}{2008}]{ive_etal_08}
{Ivezic}, Z., et~al. 2008, ArXiv e-prints

\bibitem[\protect\citeauthoryear{{Jing} \& {Suto}}{{Jing} \&
  {Suto}}{2002}]{jin_sut_02}
{Jing}, Y.~P.,  \& {Suto}, Y. 2002, \apj, 574, 538

\bibitem[\protect\citeauthoryear{{Johnston}, {Law}, \& {Majewski}}{{Johnston}
  et~al.}{2005}]{joh_etal_05}
{Johnston}, K.~V., {Law}, D.~R.,  \& {Majewski}, S.~R. 2005, \apj, 619, 800

\bibitem[\protect\citeauthoryear{{Kazantzidis} et~al.}{{Kazantzidis}
  et~al.}{2004}]{kkzanm04}
{Kazantzidis}, S., {Kravtsov}, A.~V., {Zentner}, A.~R., {Allgood}, B., {Nagai},
  D.,  \& {Moore}, B. 2004, \apjl, 611, L73

\bibitem[\protect\citeauthoryear{{Kazantzidis}, {Magorrian}, \&
  {Moore}}{{Kazantzidis} et~al.}{2004}]{kmm04}
{Kazantzidis}, S., {Magorrian}, J.,  \& {Moore}, B. 2004, \apj, 601, 37

\bibitem[\protect\citeauthoryear{{Kochanek}}{{Kochanek}}{1995}]{kochan_95}
{Kochanek}, C.~S. 1995, \apj, 445, 559

\bibitem[\protect\citeauthoryear{{Koopmans}, {de Bruyn}, \&
  {Jackson}}{{Koopmans} et~al.}{1998}]{koo_etal_98}
{Koopmans}, L.~V.~E., {de Bruyn}, A.~G.,  \& {Jackson}, N. 1998, \mnras, 295,
  534

\bibitem[\protect\citeauthoryear{{Koposov} et~al.}{{Koposov}
  et~al.}{2012}]{koposov+12}
{Koposov}, S.~E., et~al. 2012, \apj, 750, 80

\bibitem[\protect\citeauthoryear{{Kuijken} \& {Tremaine}}{{Kuijken} \&
  {Tremaine}}{1994}]{kui_tre_94}
{Kuijken}, K.,  \& {Tremaine}, S. 1994, \apj, 421, 178

\bibitem[\protect\citeauthoryear{{Lake} \& {Norman}}{{Lake} \&
  {Norman}}{1983}]{lake_norman_83}
{Lake}, G.,  \& {Norman}, C. 1983, \apj, 270, 51

\bibitem[\protect\citeauthoryear{{Laskar}}{{Laskar}}{1993}]{laskar93}
{Laskar}, J. 1993, Celestial Mechanics and Dynamical Astronomy, 56, 191

\bibitem[\protect\citeauthoryear{{Law} \& {Majewski}}{{Law} \&
  {Majewski}}{2010}]{law_majewski10}
{Law}, D.~R.,  \& {Majewski}, S.~R. 2010, \apj, 714, 229

\bibitem[\protect\citeauthoryear{{Law}, {Majewski}, \& {Johnston}}{{Law}
  et~al.}{2009}]{law+09}
{Law}, D.~R., {Majewski}, S.~R.,  \& {Johnston}, K.~V. 2009, \apjl, 703, L67

\bibitem[\protect\citeauthoryear{{Loebman} et~al.}{{Loebman}
  et~al.}{2012}]{loebman+12}
{Loebman}, S.~R., {Ivezi{\'c}}, {\v Z}., {Quinn}, T.~R., {Governato}, F.,
  {Brooks}, A.~M., {Christensen}, C.~R.,  \& {Juri{\'c}}, M. 2012, \apjl, 758,
  L23

\bibitem[\protect\citeauthoryear{{Lux} et~al.}{{Lux} et~al.}{2012}]{lux+12}
{Lux}, H., {Read}, J.~I., {Lake}, G.,  \& {Johnston}, K.~V. 2012, \mnras, L464

\bibitem[\protect\citeauthoryear{{Macci{\`o}} et~al.}{{Macci{\`o}}
  et~al.}{2007}]{mac_etal_07b}
{Macci{\`o}}, A.~V., {Dutton}, A.~A., {van den Bosch}, F.~C., {Moore}, B.,
  {Potter}, D.,  \& {Stadel}, J. 2007, \mnras, 378, 55

\bibitem[\protect\citeauthoryear{{Magnenat}}{{Magnenat}}{1982}]{magnenat_82}
{Magnenat}, P. 1982, \aap, 108, 89

\bibitem[\protect\citeauthoryear{{Martinet} \& {de Zeeuw}}{{Martinet} \& {de
  Zeeuw}}{1988}]{martinet_dezeeuw_88}
{Martinet}, L.,  \& {de Zeeuw}, T. 1988, \aap, 206, 269

\bibitem[\protect\citeauthoryear{{Mart{\'{\i}}nez-Delgado}
  et~al.}{{Mart{\'{\i}}nez-Delgado} et~al.}{2004}]{mar_etal_04}
{Mart{\'{\i}}nez-Delgado}, D., {G{\'o}mez-Flechoso}, M.~{\'A}., {Aparicio}, A.,
   \& {Carrera}, R. 2004, \apj, 601, 242

\bibitem[\protect\citeauthoryear{{Milgrom}}{{Milgrom}}{1983}]{mond}
{Milgrom}, M. 1983, \apj, 270, 365

\bibitem[\protect\citeauthoryear{{Miller} \& {Scalo}}{{Miller} \&
  {Scalo}}{1979}]{miller_scalo79}
{Miller}, G.~E.,  \& {Scalo}, J.~M. 1979, \apjs, 41, 513

\bibitem[\protect\citeauthoryear{{Moore} et~al.}{{Moore}
  et~al.}{2004}]{moo_etal_04}
{Moore}, B., {Kazantzidis}, S., {Diemand}, J.,  \& {Stadel}, J. 2004, \mnras,
  354, 522

\bibitem[\protect\citeauthoryear{{Oguri}, {Lee}, \& {Suto}}{{Oguri}
  et~al.}{2003}]{ogu_etal_03}
{Oguri}, M., {Lee}, J.,  \& {Suto}, Y. 2003, \apj, 599, 7

\bibitem[\protect\citeauthoryear{{Olling}}{{Olling}}{1995}]{olling_95}
{Olling}, R.~P. 1995, \aj, 110, 591

\bibitem[\protect\citeauthoryear{{Olling}}{{Olling}}{1996}]{olling_96}
{Olling}, R.~P. 1996, \aj, 112, 481

\bibitem[\protect\citeauthoryear{{Olling} \& {Merrifield}}{{Olling} \&
  {Merrifield}}{2000}]{oll_mer_00}
{Olling}, R.~P.,  \& {Merrifield}, M.~R. 2000, \mnras, 311, 361

\bibitem[\protect\citeauthoryear{{Pe{\~n}arrubia} et~al.}{{Pe{\~n}arrubia}
  et~al.}{2010}]{penarrubia+10}
{Pe{\~n}arrubia}, J., {Belokurov}, V., {Evans}, N.~W.,
  {Mart{\'{\i}}nez-Delgado}, D., {Gilmore}, G., {Irwin}, M.,
  {Niederste-Ostholt}, M.,  \& {Zucker}, D.~B. 2010, \mnras, 408, L26

\bibitem[\protect\citeauthoryear{{Pe{\~n}arrubia}, {Koposov}, \&
  {Walker}}{{Pe{\~n}arrubia} et~al.}{2012}]{penarrubia+12}
{Pe{\~n}arrubia}, J., {Koposov}, S.~E.,  \& {Walker}, M.~G. 2012, \apj, 760, 2

\bibitem[\protect\citeauthoryear{{Pe{\~n}arrubia} et~al.}{{Pe{\~n}arrubia}
  et~al.}{2011}]{penarrubia+11}
{Pe{\~n}arrubia}, J., et~al. 2011, \apjl, 727, L2

\bibitem[\protect\citeauthoryear{{Perryman} et~al.}{{Perryman}
  et~al.}{2001}]{gaia}
{Perryman}, M.~A.~C., et~al. 2001, \aap, 369, 339

\bibitem[\protect\citeauthoryear{{Read} \& {Moore}}{{Read} \&
  {Moore}}{2005}]{read_moore05}
{Read}, J.~I.,  \& {Moore}, B. 2005, \mnras, 361, 971

\bibitem[\protect\citeauthoryear{{Ro{\v s}kar} et~al.}{{Ro{\v s}kar}
  et~al.}{2010}]{roskar+10}
{Ro{\v s}kar}, R., {Debattista}, V.~P., {Brooks}, A.~M., {Quinn}, T.~R.,
  {Brook}, C.~B., {Governato}, F., {Dalcanton}, J.~J.,  \& {Wadsley}, J. 2010,
  \mnras, 408, 783

\bibitem[\protect\citeauthoryear{{Ro{\v s}kar}, {Debattista}, \&
  {Loebman}}{{Ro{\v s}kar} et~al.}{2013}]{roskar+13}
{Ro{\v s}kar}, R., {Debattista}, V.~P.,  \& {Loebman}, S.~R. 2013, \mnras, 433,
  976

\bibitem[\protect\citeauthoryear{{Ro{\v s}kar} et~al.}{{Ro{\v s}kar}
  et~al.}{2012}]{roskar+12}
{Ro{\v s}kar}, R., {Debattista}, V.~P., {Quinn}, T.~R.,  \& {Wadsley}, J. 2012,
  \mnras, 426, 2089

\bibitem[\protect\citeauthoryear{{Ro{\v s}kar} et~al.}{{Ro{\v s}kar}
  et~al.}{2008}]{rok_08a}
{Ro{\v s}kar}, R., {Debattista}, V.~P., {Stinson}, G.~S., {Quinn}, T.~R.,
  {Kaufmann}, T.,  \& {Wadsley}, J. 2008, \apjl, 675, L65

\bibitem[\protect\citeauthoryear{{Sackett} \& {Sparke}}{{Sackett} \&
  {Sparke}}{1990}]{sac_spa_90}
{Sackett}, P.~D.,  \& {Sparke}, L.~S. 1990, \apj, 361, 408

\bibitem[\protect\citeauthoryear{{Schoenmakers}, {Franx}, \& {de
  Zeeuw}}{{Schoenmakers} et~al.}{1997}]{sch_etal_97}
{Schoenmakers}, R.~H.~M., {Franx}, M.,  \& {de Zeeuw}, P.~T. 1997, \mnras, 292,
  349

\bibitem[\protect\citeauthoryear{{Schweizer}, {Whitmore}, \&
  {Rubin}}{{Schweizer} et~al.}{1983}]{sch_etal_83}
{Schweizer}, F., {Whitmore}, B.~C.,  \& {Rubin}, V.~C. 1983, \aj, 88, 909

\bibitem[\protect\citeauthoryear{{Sellwood}}{{Sellwood}}{2003}]{sellwo_03}
{Sellwood}, J.~A. 2003, \apj, 587, 638

\bibitem[\protect\citeauthoryear{{Slater} et~al.}{{Slater}
  et~al.}{2013}]{slater+13}
{Slater}, C.~T., et~al. 2013, \apj, 762, 6

\bibitem[\protect\citeauthoryear{{Spekkens} \& {Sellwood}}{{Spekkens} \&
  {Sellwood}}{2007}]{spekkens_sellwood07}
{Spekkens}, K.,  \& {Sellwood}, J.~A. 2007, \apj, 664, 204

\bibitem[\protect\citeauthoryear{{Stadel}}{{Stadel}}{2001}]{stadel_phd}
{Stadel}, J.~G. 2001, Ph.D.~Thesis, University of Washington

\bibitem[\protect\citeauthoryear{{Steiman-Cameron} \&
  {Durisen}}{{Steiman-Cameron} \& {Durisen}}{1984}]{steiman-cameron_durisen_84}
{Steiman-Cameron}, T.~Y.,  \& {Durisen}, R.~H. 1984, \apj, 276, 101

\bibitem[\protect\citeauthoryear{{Stinson} et~al.}{{Stinson}
  et~al.}{2006}]{sti_etal_06}
{Stinson}, G., {Seth}, A., {Katz}, N., {Wadsley}, J., {Governato}, F.,  \&
  {Quinn}, T. 2006, \mnras, 373, 1074

\bibitem[\protect\citeauthoryear{{Valluri} et~al.}{{Valluri}
  et~al.}{2010}]{valluri+10}
{Valluri}, M., {Debattista}, V.~P., {Quinn}, T.,  \& {Moore}, B. 2010, \mnras,
  403, 525

\bibitem[\protect\citeauthoryear{{Valluri} et~al.}{{Valluri}
  et~al.}{2012}]{valluri+12}
{Valluri}, M., {Debattista}, V.~P., {Quinn}, T.~R., {Ro{\v s}kar}, R.,  \&
  {Wadsley}, J. 2012, \mnras, 419, 1951

\bibitem[\protect\citeauthoryear{{Valluri} \& {Merritt}}{{Valluri} \&
  {Merritt}}{1998}]{valluri_merritt98}
{Valluri}, M.,  \& {Merritt}, D. 1998, \apj, 506, 686

\bibitem[\protect\citeauthoryear{{Wadsley}, {Stadel}, \& {Quinn}}{{Wadsley}
  et~al.}{2004}]{gasoline}
{Wadsley}, J.~W., {Stadel}, J.,  \& {Quinn}, T. 2004, New Astronomy, 9, 137

\bibitem[\protect\citeauthoryear{{Wilkinson} \& {James}}{{Wilkinson} \&
  {James}}{1982}]{wil_jam_82}
{Wilkinson}, A.,  \& {James}, R.~A. 1982, \mnras, 199, 171

\bibitem[\protect\citeauthoryear{{Zemp} et~al.}{{Zemp} et~al.}{2011}]{zemp+11}
{Zemp}, M., {Gnedin}, O.~Y., {Gnedin}, N.~Y.,  \& {Kravtsov}, A.~V. 2011,
  \apjs, 197, 30

\bibitem[\protect\citeauthoryear{{Zemp} et~al.}{{Zemp} et~al.}{2012}]{zemp+12}
{Zemp}, M., {Gnedin}, O.~Y., {Gnedin}, N.~Y.,  \& {Kravtsov}, A.~V. 2012, \apj,
  748, 54

\end{thebibliography}


\appendix

\section{An Interpretation of the Halo Instability}
\label{haloorbits}

Here we explore the cause of the halo instability which prevents discs
from inhabiting an intermediate-axis orientation.  As shown above, the
orientation of the inner potential changes as the disc is grown within
it.  In model IA1, the axes of the potential are initially ordered as
$x_\Phi > z_\Phi > y_\Phi$, but once the disc grows, the {\it inner}
potential gets flattened and has axes ordered as $x_\Phi > y_\Phi >
z_\Phi$, while at larger radii the original axes ordering is retained.
\citet{valluri+10} showed that while tube orbits are uncommon in halo
A before the disc forms, a fraction of halo box orbits are transformed
by the growing disc, with short axis tubes becoming abundant (we refer
to the axes ordering at large radii, rather than in the flattened
inner halo, to define orbit families).  Because of the radial change
in the axes ordering, particles circulating about the short axis of
the inner halo are actually IATs if they venture outside the inner
halo.  We propose that tube orbits {\it crossing} the inner halo are
destabilized by the radially varying halo orientation and drive the
instability of the inner halo.  We explore this hypothesis by
comparing models IA1 and LA1.  Model LA1, which was presented by
\citet{valluri+12}, is identical to model IA1 other than that the disc
is grown perpendicular to the long axis, which we found is a stable
orientation for this disc.  In model LA1, the original axes ordering
is $z_\Phi > x_\Phi > y_\Phi$ becoming, in the inner ($\la 15$ kpc)
halo, $x_\Phi > y_\Phi > z_\Phi$ once the disc is grown.  As with IA1,
any particles on tube orbits can be destabilized by crossing from the
flattened inner halo to larger radii.  Thus model LA1 acts as a
control in the interpretation of why IA1 (and the intermediate-axis
orientation in general) is unstable.

Orbits of dark matter particles in LA1 and IA1 were analysed using the
Laskar frequency analysis method \citep{laskar93, valluri_merritt98}
with the automated orbit classification scheme described previously
\citep{valluri+10, valluri+12}. Briefly, Laskar's method uses a
filtered Fourier transform method to obtain accurate orbital frequency
spectra from complex time series constructed from the orbital
phase-space coordinates. The frequency spectra are then decomposed
into the set of three linearly independent base frequencies (the
'fundamental frequencies') of which all other frequencies in the
spectrum are integer linear multiples. The ratios of fundamental
frequencies are rationalized following a method similar to that
described by \citet{carpintero_aguilar98}. Previously
\citep[e.g.][]{valluri+10} we only considered classification into the
traditional orbit families believed to constitute triaxial galaxies
(boxes, long-axis tubes, short-axis tubes, and various families of
resonant orbits). Here we adapted our code to also consider the
possibility that orbits may be tubes which circulate about the
intermediate axis.

We measure the degree of diffusivity of individual orbits via the
diffusion rate parameter $\log(\Delta f)$.  Since regular orbits have
fixed frequencies, a chaotic orbit can be identified if its
fundamental frequencies measured in the two consecutive time segments
change significantly \citep{laskar93}. \citet{valluri+10} showed that
even for orbits in $N$-body potentials (which are inherently noisy) it
is possible to distinguish between $N$-body jitter and true chaos via
a quantitative measurement of frequency drift by defining $\log(\Delta
f)$ as the logarithm of the change in the frequency of the leading
term in the orbit's frequency spectrum in two consecutive time
segments.  \citet{valluri+10} showed, using orbits in $N$-body
simulations of spherical haloes, that orbits with $\log(\Delta f)<
-1.2$ were regular.

We use the orbit sample described in \citet{valluri+12}: briefly, this
is a sample of orbits for $10^4$ particles in each model.  Each of
these particles was chosen at random from those within 200 kpc from
the centre before the disc was grown; the same set of particles is
used in models IA1 and LA1.  Each orbit is integrated for 50~Gyr.  The
frequency analysis is not guaranteed to produce accurate frequencies
for orbital integration times less than 20-30 orbital periods.  Table
\ref{tab:orbitfracs} lists the number of orbits of different types
with more than 30 orbital periods in our sample.  About two-thirds of
all orbits satisfy the orbital periods condition; more than half of
these are box orbits.  Model IA1 contains $\sim 30\%$ more box orbits
than model LA1.  This probably contributes to making it more unstable
since box orbits have zero average angular momentum making them easier
to tilt.

\begin{centering}
\begin{table}
\vbox{\hfil
\begin{tabular}{cccccc}\hline 
\multicolumn{1}{c}{Model} &
\multicolumn{1}{c}{Total} &
\multicolumn{1}{c}{Boxes} &
\multicolumn{1}{c}{LATs} &
\multicolumn{1}{c}{SATs} &
\multicolumn{1}{c}{IATs} \\ \hline
IA1 & 6697 & 4157 & 1400 &  378 & 762 \\ 
LA1 & 6782 & 3443 & 1316 & 2023 &   0 \\ \hline 
\end{tabular}
\hfil}
\caption{The number of orbits in the different families in the two
  models from a sample of $10^4$ orbits.  Only those orbits which complete 
  30 periods in 50 Gyr integrations are counted.}
\label{tab:orbitfracs}
\end{table}
\end{centering}

\begin{figure}
\includegraphics[angle=-90.,width=\hsize]{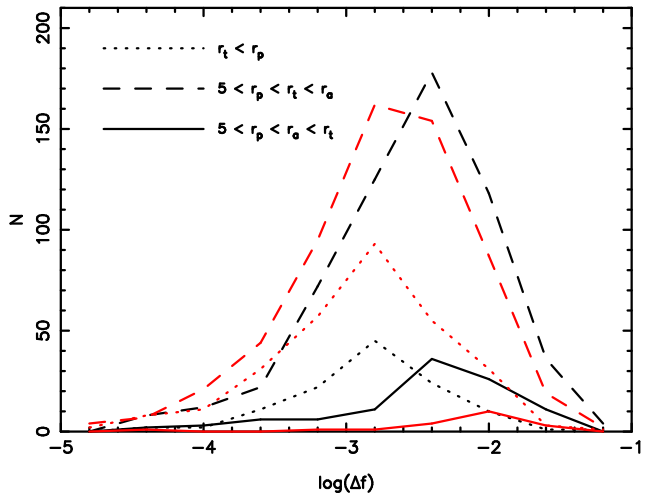}
\caption{The distributions of $\log(\Delta f)$ for tube orbits in
  model IA1 (black lines) and LA1 (red lines).  The solid, dashed and
  dotted lines show those orbits with $5 \kpc < r_{peri} < r_{apo} <
  r_t$, $5 \kpc < r_{peri} < r_t < r_{apo}$ and $r_{peri} > r_t$,
  respectively.  For model IA1, $r_t = 25$ kpc while for LA1 $r_t =
  15$ kpc.}
\label{fig:diffusion}
\end{figure}

Figure \ref{fig:diffusion} plots the distribution of $\log(\Delta f)$
for tube orbits of all types in models IA1 and LA1, separated into
three groups by radial range: $5 \kpc < r_{peri} < r_{apo} < r_{\rm
  t}$, $5 \kpc < r_{peri} < r_{\rm t} < r_{apo}$, and $r_{\rm t} <
r_{peri}$, where $r_{peri}$ and $r_{apo}$ are the peri- and apocentre
distances and $r_{\rm t}$ is the radius at which the potential
switches orientation.  From Figure \ref{fig:potshapes} we find $r_{\rm
  t} = 25 \kpc$ for model IA1, whereas a similar measurement for LA1
gives $r_{\rm t} = 15 \kpc$.  Orbits that never visit the inner region
have low $\log(\Delta f)$.  Orbits that remain wholly within the inner
region have higher diffusion rates, but they tend to be less numerous.
Orbits that move across $r_{\rm t}$ are the most abundant and have
higher diffusivity in model IA1 than in LA1.  It is this difference in
the diffusion of tube orbits crossing the radius at which the
potential reorients that we propose accounts for the different
stability properties of models IA1 and LA1.

\label{lastpage}

\end{document}